\documentclass[traditabstract]{aa}

\usepackage{txfonts}
\usepackage{natbib}
\usepackage{ragged2e}
\usepackage{graphics,graphicx,epsfig}
\usepackage{booktabs}
\usepackage{ltxtable}
\usepackage{amssymb}
\usepackage{wasysym} 

\begin{document}

\newcommand{\lele}[3]{{#1}\,$\le$\,{#2}\,$\le$\,{#3}}
\newcommand{\tauav}[1]{$\tau_{#1}$/A$_{\rm V}$}
\newcommand{\iav}[1]{I$_{#1}$/A$_{\rm V}$}
\newcommand{\fblue}[1]{F$_{70}$\,}
\newcommand{\fgreen}[1]{F$_{100}$\,}
\newcommand{\fred}[1]{F$_{160}$\,}
\def\absmag{$H_\mathrm{V}$}
\def\tss{$T_\mathrm{ss}$}
\def\rh{$r_\mathrm{h}$}
\def\espec{$\epsilon_\nu$}
\def\ebol{$\epsilon_\mathrm{bol}$}
\def\deq{$D_\mathrm{eq}$}
\def\geomalb{$p_\mathrm{V}$}
\def\qg{2001\,QG$_{298}$}
\def\deff{D$_\mathrm{eff}$}
\def\gcc{g\,cm$^{-3}$}
\def\tiunit{$\mathrm J\,m^{-2}\,s^{-1/2}K^{-1}$}

\sloppy

\title{TNOs are Cool! A Survey of the transneptunian Region} 
\subtitle{XV. Physical characteristics of 23 resonant transneptunian \\ and scattered disk objects}
%
%
\author{ A.~Farkas-Tak\'acs\inst{1,2}
\and Cs.~Kiss\inst{1,3} 
\and E.~Vilenius\inst{4}
\and G.~Marton\inst{1,3}
\and T.G.~M\"uller\inst{5}
\and M.~Mommert\inst{6}
\and \\
J.~Stansberry\inst{7}
\and E.~Lellouch\inst{8}
\and P.~Lacerda\inst{9}
\and A.~P\'al\inst{1,2}}
\institute{Konkoly Observatory, Research Centre for Astronomy and Earth Sciences, 
        Hungarian Academy of Sciences, 
    Konkoly Thege 15-17, H-1121~Budapest, Hungary
    \and 
    E\"otv\"os University, Faculty of Science, P\'azm\'any P. s\'et\'any 1/A, H-1171, Budapest, Hungary
        \and
        ELTE E\"otv\"os Lor\'and University, Institute of Phyiscs, P\'azm\'any P. s\'et\'any 1/A, H-1171, Budapest, Hungary
        \and
        Max-Planck-Institut f\"ur Sonnensystemforschung, Justus-von-Liebig-Weg 3, 37077 G\"ottingen, Germany
        \and 
        Max-Planck-Institut f\"ur extraterrestrische Physik, Giessenbachstrasse,
    85748 Garching, Germany
    \and 
    Lowell Observatory, 1400 W. Mars Hill Rd., Flagstaff, AZ, 86001, USA
    \and
    Space Telescope Science Institute, 3700 San Martin Dr., Baltimore, MD 21218, USA 
    \and
    Observatoire de Paris, Laboratoire d'\'Etudes Spatiales et
    d'Instrumentation en Astrophysique (LESIA),
    5 Place Jules Janssen, 92195 Meudon Cedex, France
    \and
    Astrophysics Research Centre, School of Mathematics and Physics, Queen's University Belfast, Belfast, UK}

\date{\today / \today}

\abstract{
The goal of this work is to determine the physical characteristics of resonant, detached and scattered disk objects in the trans-Neptunian region, observed mainly in the framework of the ``TNOs are Cool'' Herschel Open Time Key Programme. Based on thermal emission measurements with the Herschel/PACS and Spitzer/MIPS instruments, we determine size, albedo, and surface thermal properties for 23 objects using radiometric modeling  techniques. This is the first analysis in which the physical properties of objects in the outer resonances are determined for a notable sample. In addition to the results for individual objects, we  compared these characteristics with the bulk properties of other populations of the trans-Neptunian region. The newly analyzed objects show a large variety of beaming factors, indicating a diversity of thermal properties, and in general they follow the albedo-color  clustering identified earlier for Kuiper belt objects and Centaurs, further strengthening the evidence for a compositional discontinuity in the young Solar System. }


\titlerunning{``TNOs are Cool'' XV: 23 resonant trans-Neptunian and scattered disc objects}
\authorrunning{A. Farkas-Tak\'acs et al.} 

\maketitle



\section{Introduction \label{introduction}}

Trans-Neptunian objects in mean motion resonance with Neptune \citep[RTNOs; see, e.g.,][and references therein]{Gladman2012} show a dynamical behavior that their resonance angle $\phi_{jk}$ = j$\lambda$ - k$\lambda_N$ - (j-k)$\varpi$ (where $\lambda$ is the longitude of object, $\lambda_N$ the longitude of Neptune, and $\varpi$ the longitude of perihelion of the RTNO, for a specific j:k resonance) cannot have all values between 0\degr{} and 360\degr.
Typically, the resonance angle is confined to a mean value with some relatively small libration amplitude.
An important feature of the resonances is that they may provide  protection against perturbations and allow large eccentricity orbits to survive for the age of the Solar System.
Even so, the chaotic nature of the resonance border allows temporary trapping of objects near the border of the resonance, and nearly resonant objects can escape into a dynamical regime where perturbations may direct them out of the Kuiper belt toward the inner Solar System.
Resonant objects also serve as a diagnostic tool of the planetary migration era. 
Current population estimates of RTNOs are consistent with a scenario that they were likely put in place in the resonances during the planet migration era at the end of the giant-planet formation process \citep{Gladman2012}.
RTNOs show dynamical characteristics (e.g., inclination distribution) different from those of the classical Kuiper belt, but may be connected to the scattered disk population \citep{Gomes2008}, objects on high eccentricity orbits with perihelia beyond Neptune, and semi-major axes beyond the 2:1 resonance.
Recent studies \citep{Yu2018} indicate that a significant fraction of all scattered objects are transiently stuck in mean motion resonances, suggesting that these objects  originated from the same single population. 

The Outer Solar System Origins Survey \citep[OSSOS][]{Bannister} recently provided a significant increase in the number of RTNOs, adding 313 new objects of which 132 are plutinos. This survey also detected two objects in the distant 9:1 resonance \citep{Volk} that may have originated in the scattered population, and that became trapped in this resonance within the last $\sim$1\,Gyr, but could also be remnants of a larger, primordial population. 

In addition to dynamical properties \citet{Sheppard2012} 
obtained broad-band color of 58 resonant objects, with significantly different color distributions in the different resonances. For example,  those in the 5:3 and 7:4 resonances are dominated by ultra-red material, similar to that in the cold classical belt. Objects in the inner 4:3 and distant 5:2 resonances show mostly moderate red colors, similar to the scattered and detached populations, while the 2:1 and 3:2 resonances show a wide range of colors. Apart from the dynamical properties,  color, albedo, and size are very crucial physical characteristics of these objects; however, they can be determined only when thermal emission data or occultation measurements are available. 


In the framework of the ``TNOs are Cool'': A survey of the trans-Neptunian region, a Herschel Open Time Key Programme \citep{Muller2009}, several papers have been published on the physical characteristics of Centaurs and trans-Neptunian objects in population-specific papers, and  in other papers concentrating on specifically selected objects \citep{Pal12, For13} or on the general properties of these small bodies \citep{Lellouch2013}. These papers used thermal emission data obtained with the PACS photometer camera of the Herschel Space Observatory in the far-infrared, at 70, 100, and 160\,$\mu$m. The population-specific papers included the Centaurs \citep{Duffard14}, classical Kuiper belt objects \citep{Vilenius12,Vilenius14}, plutinos \citep{Mommert12},  some scattered disk and detached objects \citep{SS12}, and also the members of the Haumea collisional family \citep{Vilenius2018}. 

Using these data, in particular the size and albedo derived using radiometric models that were supplemented by color information, \citet{Lacerda2014} found evidence for a compositional discontinuity in the early Solar System, identifying two distinct types of planetary bodies: objects with dark (geometric albedo of p$_V$\,$\approx$\,0.05) and neutrally colored surfaces, or with bright (p$_V$\,$\approx$\,0.15) and red surfaces. Objects in relic populations (cold classicals, detached objects, and outer resonances, i.e.,  the 2:1 mean-motion resonance and beyond) show exclusively bright-red surfaces, while objects in dynamically less stable populations (Centaurs, plutinos, scattered disk object, and hot classicals) contain a mixture of bright-red and dark-neutral surfaces.

The largest set of RTNOs with reliable size and albedo information available are the plutinos. \citet{Mommert12} derived these physical characteristics for 18 objects as part of the ``TNOs Are Cool'' program. They found sizes ranging from 150 to 730\,km and geometric albedos varying between 0.04 and 0.28, with an average albedo of 0.08$\pm$0.03, similar to the mean albedo of Centaurs, Jupiter family comets, and other TNOs except cold classicals and detached objects. Based on these results, the cumulative power law of the size distribution in this dynamical group is q\,=\,2 between sizes of 120 to 400\,km, and q\,=\,3 for larger objects. 

In a similar study \citep{SS12} investigated 15 scattered disk and detached objects and found notably different albedos between the scattered disk and detached subpopulations (mean geometric albedos of 6.9\% and 17\%, respectively) . 
 
Data from both studies were incorporated in the paper by \citet{Lacerda2014}, with additional objects from the incomplete Herschel sample of RTNOs in non-3:2 resonances and additional scattered disk objects from \citet{Lellouch2013}. A recent summary of thermal infrared observations of Centaurs and trans-Neptunian objects can be found in \cite{Muller2019}. 

In this paper we present the results of an investigation similar to those in the ``TNOs are Cool'' sample papers, focusing on Herschel/PACS observations of as-yet unpublished resonant and scattered disk objects. We derive size and albedo from Herschel/PACS measurements and derive radiometric models of 15 resonant, 7 scattered disk,  and 1 detached object, in some cases supplemented by data from the MIPS camera of the Spitzer Space Telescope, not presented in earlier papers. We also reconsidered the data of a scattered disk and a detached object with inconsistent flux densities or problematic radiometric fits in previous papers, derived new flux densities using the final version of our data reduction pipeline, and obtained updated radiometric size and albedo estimates. Our results are then compared with color-albedo data of the trans-Neptunian dynamical populations derived in previous papers.  

\section{Observations, data reduction, and radiometric modeling}

\subsection{Herschel-PACS data}

All Herschel-PACS measurements presented in this paper as new data were taken in scan-map mode with the PACS photometer arrays, following the observation strategy of the ``TNOs are Cool'' Open Time Key Programme \citep{Muller2009,Vilenius12,Kiss2014}. Reduction of PACS photometer data is performed using the latest version of the ``TNOs are Cool'' pipeline. Both the key program observation strategies and the main characteristics and features of the pipeline are summarized in \citet{Kiss2014}. 

All Herschel/PACS measurements presented here as new reduction followed the same observation strategy: the target is observed at two epochs, typically separated by  a day, and the data of the two epochs are combined to allow for  optimal background elimination. At each epoch we observed the target with two 70/160\,$\mu$m and two 100/160\,$\mu$m filter astronomical observation requests (AORs), corresponding to 70\degr{} and 110\degr{} scan direction orientations. Altogether we obtained eight AORs per target, and correspondingly eight individual observation idendifiers (OBSIDs; see Table~\ref{tab:pacsdata}). Then the data of these OBSIDs are combined to obtain the final products. We note that 2003\,UY$_{117}$ was measured with Herschel/PACS using the 70/160\,$\mu$m filter combination only, while all other targets were observed with both filter combinations.

We used the following main settings to obtain Level-2 flux-calibrated PACS scan maps from the raw PACS measurements:

\begin{itemize}
\item Scan legs are extended based on the slew speed 15 and 25\arcsec\,$s^{-1}$, around the commanded 20\arcsec\,$s^{-1}$ scan speed. 
\item High pass filtering with filter width of 8, 9, and 16 is used 
at 70, 100, and 160\,$\mu$m, respectively  
(high pass filter width sets the number of frames [2n+1] used
for median subtraction from the detector timeline; see Popesso et al. 2012 and Balog et al. 2014 for a detailed description of the method).
\item Pixel masking is used  above 2$\sigma$, and at the source position with 2$\times$FWHM radius
\item  Second-level deglitching is applied with nsigma\,=\,30, the sigma-clipping parameter of this deglitching method working on the map level (see the \citet{PACS_guide} for more details). 
\item Correction for the apparent motion of the target is not applied to allow an optimal evaluation of the sky background using multiple measurements.   
\item  The drizzle method is used to project the time-line data and to produce the single maps using the {\it photProject()} task in HIPE, with a pixel fraction parameter of 1.0. 
\end{itemize}

As a standard setting for all ``TNOs are Cool'' data products we use pixel sizes of 1\farcs1, 1\farcs4, and 2\farcs1 in the PACS 70, 100, and 160\,$\mu$m bands, different from the ``physical'' pixel sizes of 3.2\arcsec\ and 6.4\arcsec\ for the blue and red detectors, respectively. This allows an optimal sampling of the respective point spread functions \citep[see][]{Kiss2014}. 
Our flux calibration is based on a set of standard stars with the same basic data reduction settings as in the case of our Solar System targets.
Determination of the photometric uncertainty in scan maps is performed using the implanted source method, as described in \citet{Kiss2014}. 
The flux densities of Solar System targets are derived from the double-differential products \citep[see][for details]{Kiss2014}. 
These double-differential images, as well as co-added and simple differential images, are available in the Herschel Science Archive\footnote{http://archives.esac.esa.int/hsa/whsa} as User Provided Data Products (UPDPs), along with a detailed description of the processing steps and the data products (UPDP Release Note Version 1.0, May 5, Kiss et al., 2017). 
The summary of the observations, as well as the derived  in-band flux densities, are measured with the specific instrument--filter combinations) are presented in Table~\ref{tab:pacsdata}.

 We   reanalyzed the PACS observations of two targets from \citet{SS12}, 1999\,KR$_{16}$ and 2005\,QU$_{182}$. In both cases in the earlier evaluation the 100\,$\mu$m flux densities were inconsistently low compared with the 70 and 160\,$\mu$m flux densities.  While there are some differences between the \citet{SS12} and the present calibration and data reduction (e.g., different HIPE version, absolute calibration) the main difference is that in \citet{SS12}  a different kind of data product, namely supersky-subtracted images were used for final photometry. As  discussed in \citet{Kiss2014,Kiss2017}, double-differential (DDIFF) maps are superior in obtaining accurate and reliable photometry compared with supersky-subtracted maps, and we used these DDIFF maps for 1999\,KR$_{16}$ and 2005\,QU$_{182}$, reduced with the latest version of our pipeline \citep{Kiss2017}. The supersky-subtracted images are more sensitive to the presence of background sources close to the target at one of the epochs considered. This was the case for the PACS measurements of both targets. By using these DDIFF images  the 100\,$\mu$m flux densities are now consistent with those in the two other bands, as confirmed by the subsequent radiometric modeling (see Sect.~\ref{sect:results}).
 
In addition to these objects, updated flux densities of 2007\,UK$_{126}$, along with albedo and size derived from a combination of radiometric modeling and occultation data, are available in \citet{Schindler2017}. Similarly, a revised radiometric model is available for 2007\,OR$_{10}$ in \citet{Kiss2019}. All Herschel/PACS data are available in the ``Small Bodies: Near and Far'' database of thermal infrared observations of Solar System small bodies\footnote{https://ird.konkoly.hu}  \citep{Kiss2019,Szakats2019}. 

\begin{table*}
\caption{Herschel in-band fluxes at all three PACS bands for the ``new'' objects, and for two objects with updated fluxes, presented in this work. Column headings (from left to right): object name; Dynamical classification (Res: resonant, SDO: scattered disk object, DO: detached object); OBSID:  observation identifiers in the Herschel Science Archive; t$_{\mathtt{OBS}}$:  total duration of the two visits; JD: measurement mid-time in Julian date; r$_h$: heliocentric distance; $\Delta$: observer-to-target distance; $\alpha$: phase angle; F$_{70}$, F$_{100}$, and F$_{160}$:  in-band flux densities obtained in the PACS 70, 100, and 160\,$\mu$m bands, respectively.  Upper limits correspond to 2$\sigma$ uncertainties derived with the implanted source method \citep{Vilenius14,Kiss2014}. 
*: target from \cite{SS12} and **: target from \cite{Vilenius2018} reanalyzed with the latest version of our data reduction pipeline.}
\centering
\scriptsize
\begin{tabular}{cccccccc|ccc}
\hline
Object  & Dyn. & OBSID & t$_{\mathtt{OBS}}$ &  JD & r$_h$    & $\Delta$ & $\alpha$ & F$_{70}$ & F$_{100}$ & F$_{160}$ \\
        & cl.   &     &  (sec)         & (day) & (AU) & (AU)     &  (deg) &   (mJy)  & (mJy)   & (mJy)  \\
\hline
\hline
(26181) 1996\,GQ$_{21}$ & SDO &1342212818/...821 & 5656  & 2455579.43 & 41.6056 & 41.9751 & 1.26 & 4.51$\pm$1.01        &       6.78$\pm$1.28   &       7.53$\pm$2.15  \\
                          & &1342213075/...078 & 5656  & 2455580.23 & 41.6063 & 41.9628 & 1.26 & & & \\
(523588) 2000\,CN$_{105}$ & Res 9:5 &1342197691/...694 & 3400 & 2455351.20 & 46.5639 & 46.6134 &1.25  &  $<$ 1.31  &       $<$ 1.50        &       $<$ 1.63  \\        
                            & &1342197781/...784 &3400 & 2455352.41 & 46.5641 & 46.6340 & 1.25 &  \\
(82075) 2000\,YW$_{134}$ & Res 8:3 & 1342196008/...011 & 4528  & 2455325.61 & 44.2099 & 44.3812 &1.30  & 4.88$\pm$1.07      & 5.13$\pm$1.47 &       4.88$\pm$2.13  \\
                            & & 1342196133/...036 & 4528  & 2455326.83 & 44.2105 & 44.4024 & 1.29 &  \\
                            & & 1342187074 & 5666 & 2455154.17 & 44.1263 & 43.8338 & 1.25 &  & -- & \\
(82155) 2001\,FZ$_{173}$ & SDO & 1342236630/...633 & 4528 & 2455933.26 & 32.4131 & 32.5753 & 1.72  &  8.51$\pm$1.54 & 7.91$\pm$1.77      &       5.01$\pm$2.46  \\     
                           & & 1342236908/...011 & 4528 & 2455934.17 & 32.4132 & 32.5596 & 1.73 &\\
(139775) 2001\,QG$_{298}$ & Plutino & 1342213211/...214&5656  & 2455585.66  & 31.7619 & 32.1381 & 1.64  & $<$ 0.913        &       $<$ 1.19        &       $<$ 2.00  \\        
                            & & 1342213266/...269&5656  &  2455586.32 & 31.7619 & 32.1489 & 1.64 &  \\
2001\,QX$_{322}$ & SDO & 1342211619/...622 & 5656  & 2455557.80 & 41.3329 & 41.0520 & 1.32  & $<$ 0.99    &       $<$ 1.06        &       $<$ 1.51  \\     
                            & & 1342211807/...010 & 5656  & 2455558.58 & 41.3336 & 41.0646 & 1.32 &  \\
(42301) 2001\,UR$_{163}$ & SDO & 1342199507/...510 & 2272  & 2455378.59 & 50.9591 & 51.3267 & 1.07 &  2.88$\pm$1.37       &       $<$1.69 &       $<$3.09  \\     
                            & & 1342199650/...653 & 2272  & 2455380.07 & 50.9601 & 51.3042 & 1.08 &  \\
(126154) 2001\,YH$_{140}$ & Res 5:3 & 1342206036/...039 & 3400 & 2455477.34 & 36.6150 & 36.9327 & 1.49 & 5.08$\pm$1.40 & 3.67$\pm$1.70 & $<$1.95 \\
                    && 1342206056/...059 & 3400 & 2455478.10 & 36.6151 & 36.9203 & 1.49 & \\
                    & & 1342187062 & 5666 & 2455153.28 & 36.5768 & 36.1686 & 1.44 & & -- &  \\
(119878) 2002\,CY$_{224}$ & Res 12:5 & 1342195506/...509 & 3400  & 2455311.51 & 37.2072 & 36.8614 & 1.47  & $<$ 1.59  &       $<$ 1.78        &       $<$ 2.1  \\
                            & & 1342195610/...613 & 3400  & 2455313.49 & 37.2082 & 36.8945 & 1.49 & \\
2002\,GP$_{32}$  & Res 5:2 &  1342204144/...147 &3400  & 2455448.42 & 32.1569&32.3768&1.76 & 3.09$\pm$0.95 &       5.43$\pm$1.53   &       3.58$\pm$1.29  \\       
                           & &  1342204204/...207 &3400  & 2455449.05 & 32.1570&32.3875&1.75 &  \\
(133067) 2003\,FB$_{128}$ & Plutino & 1342237146/...149 &4528  & 2455938.19& 34.2390&34.6867&1.47 & 3.22$\pm$0.90    &       3.80$\pm$1.44   &       $<$ 1.94  \\                
                            & & 1342237226/...229 &4528  & 2455938.85 & 34.2395&34.6769&1.47 &  \\
(469505) 2003\,FE$_{128}$ & Res 2:1 & 1342237150/...153 & 5656  & 2455938.25 & 35.8635 & 36.3139 & 1.40  &  3.08$\pm$1.00    &       $<$ 1.39        &       $<$ 2.51  \\        
                           & & 1342237230/...033 & 5656 & 2455938.91  & 35.8635 & 36.3037 & 1.41 &  \\
(143707) 2003\,UY$_{117}$  & Res 5:2 & 1342238745/...746 &3392  & 2455965.35 & 32.8901&33.1340&1.67 &  6.13$\pm$0.76 &       --      &       $<$ 2.16  \\     
                             & & 1342238790/...791 &3392  & 2455965.98 &32.8903&33.1447&1.66  &  \\
(455502) 2003\,UZ$_{413}$ & Plutino &1342212760/...763 &2272 & 2455578.11 & 42.6153&42.2946&1.26 & 21.49$\pm$1.45 &       17.02$\pm$1.71  &       14.82$\pm$3.53  \\ 
                           & &1342212858/...861 &2272 & 2455579.85 & 42.6162&42.3239&1.27 &  \\
(450265) 2003\,WU$_{172}$ & Plutino & 1342250794/...797 & 2836  & 2456180.54 & 29.6443&29.8432&1.91 & 6.29$\pm$1.40  &       9.44$\pm$1.38   &       3.92$\pm$1.87  \\     
                             & & 1342250830/...833 & 2836  & 2456181.36 & 29.6441&29.8292&1.92 &  \\
(175113) 2004\,PG$_{115}$ & SDO & 1342219009/...012 & 2256 & 2455670.53 & 36.9594 & 37.3635 & 1.43 &  6.68$\pm$2.39       &       8.43$\pm$3.03   &       $<$ 3.15  \\
                                & & 1342219048/...051 & 2256 & 2455671.09 & 36.9596 & 37.3554 & 1.44 & \\
(145451) 2005\,RM$_{43}$ & SDO & 1342202281/...284 & 2272 & 2455417.29 & 35.4298 & 35.6640 & 1.60 & 15.20$\pm$1.18       &       16.55$\pm$2.87  &       14.79$\pm$2.41   \\
                          & & 1342202320/...323 & 2272 & 2455417.99 & 35.4301 & 35.6527 & 1.60 &\\
(308379) 2005\,RS$_{43}$ & Res 2:1 & 1342213502/...505 & 4528 & 2455592.24 & 42.4029 & 42.7677 & 1.24 &3.09$\pm$0.94 & 3.45$\pm$1.46 & $<$1.83 \\
                    && 1342213558/...561 & 4528 & 2455593.00 & 42.4033 & 42.7805 & 1.24 & \\    
\hline
(40314) 1999\,KR$_{16}$** & DO &1342212814/...817 & 5844 & 2455579.36 & 35.7593 & 36.0633 & 1.51 & 7.61$\pm$1.44 & 5.21$\pm$2.29 & $<$ 1.98  \\
                                && 1342213071/...074 & 5844 & 2455580.16 & 35.7589 & 36.0497 & 1.51 & \\
(303775) 2005\,QU$_{182}$* & SDO & 1342212619/...622 & 2460 & 2455576.11 & 48.8994 & 49.1349 & 1.13 & 4.41$\pm$1.28      &       6.94$\pm$2.11   &       $<$ 2.73  \\
                                && 1342212696/...699 & 2460 & 2455577.08 & 48.9008 & 49.1522 & 1.12 & \\
\hline
\end{tabular}
\label{tab:pacsdata}
\end{table*}

\subsection{Spitzer/MIPS data}

Spitzer/MIPS measurements were reduced using the same pipeline as was used for the reduction of MIPS data of Centaurs and trans-Neptunian objects by 
\citet{Migo} and \citet{stansberry2008,stansberry2012}. 
The MIPS instrument team data analysis tools \citep{Gordon2005} were used to produce flux-calibrated images for each band, and the contribution of background objects were subtracted (see \citealt{stansberry2008}). 
Aperture photometry was performed  on the original images and on the final images, and the final flux values were obtained using the aperture corrections by \citet{Gordon2007} and \citet{Engelbracht2007}. 
The flux densities obtained are presented in Table~\ref{tab:spitzer}. 
In four cases Spitzer/MIPS flux densities were obtained in addition to the PACS fluxes (upper five rows in Table~\ref{tab:spitzer}). 
We have three targets with Spitzer data only. In these cases the same methods were applied as in the case of PACS only and PACS/MIPS combined measurements.
Spitzer-only targets were the plutinos 1995\,QY$_9$ and 2003\,QX$_{111}$ and the Neptune trojan 2001\,QR$_{322}$.  

\begin{table*}
\caption{Spitzer observations summary. Columns headings (from left to right): objects name; dynamical class; Spitzer observation identifier; observation date; heliocentric distance; target-to-observer distance; phase angle; Spitzer/MIPS 24\,$\mu$m in-band flux density; Spitzer/MIPS 71\,$\mu$m in-band} flux density.
\scriptsize
\centering
\begin{tabular}{ccccccc|cc}
\hline
object & Dyn. & AORkeys & JD & r$_h$ & $\Delta$ & $\alpha$ & F$_{24}$ & F$_{71}$ \\
     &    cl.    & (day) & (AU) & (AU) & (deg) & (mJy) & (mJy) \\
\hline 
\hline
1996\,GQ$_{21}$ & SDO & 9038848 & 2453222.29 & 39.937 & 39.977 & 1.46 & --- & $<$ 3.537 \\
2000\,YW$_{134}$ & Res 8:3 & 25234688, 944, 25235200 & 2454829.68 & 43.967& 43.604 & 1.24 & 0.032 $\pm$ 0.015 & $<$ 1.29 \\
                 &         & 9039872 & 2453108.12 & 43.197 & 42.988 & 1.30 & --- & 4.676 $\pm$ 2.186 \\
2001\,QX$_{322}$ & SDO  & 9036544 & 2453368.70 & 39.617 & 39.393 & 1.42 & --- & $<$ 3.08 \\
2001\,UR$_{163}$ & SDO  & 9031168 & 2453365.94 & 49.554 & 49.099 & 1.04 & --- & $<$ 2.474 \\
2001\,YH$_{140}$ & Res 5:3 & 17770240, 496 & 2454046.21 & 36.477 & 36.525 & 1.60 & $<$ 0.052 & 5.602 $\pm$ 1.40 \\
1999\,KR$_{16}$  & DO   & 9036288 & 2453784.24 & 36.728 & 36.649 & 1.56 & --- & $<$ 0.712\\
\hline
\hline
1995\,QY$_{9}$ & Plutino & 9034752 & 2453177.92 & 29.230 & 29.343 & 1.98 & $<$ 0.127 & $<$6.896 \\
2001\,QR$_{322}$ & 1:1 & 11090176/0432/688/944 & 2453368.27 & 29.708 & 29.623 & 1.94 & 0.129 $\pm$ 0.018 & 3.704 $\pm$ 0.644 \\
2003\,QX$_{111}$ & Plutino & 11099392/648/904,11100160/416/672 & 2453366.08 & 39.476 & 39.472 & 1.46 & 0.032 $\pm$ 0.009 & 2.729 $\pm$ 0.603 \\
\hline
\end{tabular}
\label{tab:spitzer}
\end{table*}

\subsection{Radiometric modeling \label{sect:radmod}}

In modeling the thermal emission of our targets we followed the main steps presented in the previous ``TNOs are Cool'' population-specific papers, in particular those used in \citet{Vilenius14}. 
We use the Near-Earth Asteroid Thermal Model (NEATM; \citealt{Harris98}) to obtain the temperature distribution on the surface of a body that is assumed to be airless, spherical, and in instantaneous equilibrium with the solar radiation. 
Deviations from the temperature distribution of a smooth surface and non-rotating body are considered through the beaming parameter $\eta$ which reflects the combined effects of spin properties, thermal inertia and surface roughness \citep[see, e.g.,][]{Spencer1989,Spencer1990}.  
 With the application of the beaming parameter the temperature of the subsolar point $T_{ss}$ is calculated as
\begin{equation}
    T_{ss} = {\Big({{S_{\odot}(1-p_Vq)}\over{\epsilon\sigma\eta r_h^2}}\Big)^{1\over4}}
,\end{equation}
\noindent where $S_{\odot}$ is the solar irradiation at 1\,AU, $q$ the phase integral, $\epsilon$ the bolometric emissivity, and r$_h$ the heliocentric distance of the target. The temperature distribution on the surface is obtained as 
\begin{equation}
    T(\vartheta) = T_{ss}\,cos^{1/4}\vartheta
,\end{equation}
\noindent where $\vartheta$ is the angular distance form the subsolar point on the surface of the body. 

We use a constant  spectral emissivity of $\epsilon(\lambda)$\,=\,0.9 throughout the wavelength range of our measurements (24--160\,$\mu$m),  hence a bolometric emissivity of $\epsilon$\,=\,0.9 in the calculation of T$_{ss}$. While the spectral emissivity has been shown to decrease for longer submillimeter wavelengths, its value is fairly constant below $\sim$200\,$\mu$m \citep{For13,Lellouch2017}.  We fit the free parameters of the NEATM model of a target (geometric albedo, \geomalb; effective diameter, \deff\, of an equal area sphere; and beaming parameter, $\eta$) by minimizing the reduced-$\chi^2$ values, calculated from the observed and modeled flux densities and the observed uncertainties. 

Our NEATM model calculates in-band flux densities for our instrument--filter combinations in addition to the monochromatic flux densities that a NEATM model calculates by default. As the local blackbody temperature is known at each grid point of this computation (using the standard $T_{ss} \cos^{1/4}\vartheta$ assumption) it is possible to calculate the actual color-correction factors that have to be applied to obtain the in-band flux densities from the monochromatic flux densities \citep{Muller2011,Stansberry2007}. These in-band flux densities are integrated (summed) to obtain the total, disk-integrated in-band flux density of the target for that specific model and instrument--filter. In the $\chi^2$ minimization we use these in-band model flux densities with the measured in-band flux densities, instead of using color-corrected monochromatic values, as  was done in previous similar studies \citep[e.g.,][]{Mommert12,SS12,Vilenius14}. This eliminates the uncertainty in the determination of the color correction from measured in-band flux densities to monochromatic ones. The typical $C_\lambda$ 
color-correction factors (used to transform the $F'$ monochromatic flux densities to the $F$ in-band flux densities as
$F_{\lambda}$\,=\,$F_\lambda' \cdot C_\lambda$) are  $C_{70}$\,=\,0.98--1.03, $C_{100}$\,=\,0.98--0.99, and $C_{160}$\,=\,0.98--1.01 for the PACS band, and $C_{24}$\,=\,1.00--1.15 and $C_{71}$\,=\,0.89--0.97 for the MIPS bands. 
In some cases MIPS  color-correction factors are significant (especially in the 24\,$\mu$m band, due to the relatively low surface temperatures of the targets) resulting in a $\sim$10\% change between the in-band and monochromatic flux densities, in contrast to the PACS measurements for which the change in flux density is $\leq$3\% in all bands.

We consider the absolute magnitude H$_V$ of the object as a measurement constraining the relationship between the geometric albedo and the effective diameter, reducing the degrees of freedom. 
H$_V$ values were mostly taken from the literature (see Table~\ref{table:colours}), but for 2003\,WU$_{172}$ it was obtained from the data in the Minor Planet Center database .  

The beaming parameter $\eta$ is a complex function of the basic characteristics of the targets, including the spin-axis orientation via the corresponding  subsolar latitude $\beta_{ss}$, the thermal parameter $\Theta$, and the surface roughness, described by the root mean square surface slopes $s$. The thermal parameter $\Theta$ is defined as
\begin{equation}
   \Theta = {{\Gamma\sqrt{\omega}}\over{\epsilon\sigma T_{ss1}^3}}   
,\end{equation}
\noindent where $\Gamma$ is the thermal inertia, $\omega$ is the spin rate of the target, and T$_{ss1}$ is the subsolar temperature uncorrected for beaming ($\eta$\,$\equiv$\,1):
\begin{equation}
    T_{ss1} = {\Big({{S_{\odot}(1-p_Vq)}\over{\epsilon\sigma r_h^2}}\Big)^{1\over4}}
\end{equation}

The dependence of $\eta$ on these parameters is discussed in detail in \citet{Lellouch2013}, based on \citet{Spencer1989} and \citet{Spencer1990}. In this approach $\eta$\,$>$\,1 is caused by a combined effect of thermal inertia and rotation, and scales with the thermal parameter $\Theta$ and the subsolar latitude $\beta_{ss}$ in the ``no roughness'' case; $\eta$\,$<$\,1 is explained by surface roughness effects. We used the Brucker et al. (2009) formula to obtain a geometric albedo-dependent phase integral in the calculation of the Bond albedo, $A$\,=\,$1-p_Vq$. 

In some cases our fits provided solutions with large error bars or converged to our lower and upper beaming parameter limits, $\eta$\,=\,0.5 and 2.5. The lower limit of $\eta$\,=\,0.5 was set considering the possible highest surface roughness correction in the beaming parameter for very low $\Theta$ values \citep{Spencer1990,Lellouch2013}. Similarly, the upper limit of $\eta$\,=\,2.5 is set according to the maximum possible values for $\beta_{ss}$\,=\,0\degr\, subsolar latitude, T$_{ss}$\,=\,60-70\,K (typical subsolar temperatures of our ``high $\eta$'' targets), and high $\Theta$\,$\geq$\,100 thermal parameter values \citep{Spencer1990,Lellouch2013}. In these cases we repeated the analysis with a ``fixed'' $\eta$ of 1.25$\pm$0.35, using this beaming parameter range to estimate the errors in \geomalb\, and \deff\, \citep[see also][]{Vilenius14}. These cases are also listed in Table~\ref{table:results}. This fixed beaming parameter value is slightly higher than the $\eta$\,=\,1.2 used by \citet{stansberry2008} and \citet{Lellouch2013} as in our sample the beaming parameters obtained from acceptable fits provided a higher average value (see Sect.~\ref{sect:generalresults}). Flux density upper limits were treated in the same way as described in \citet{Vilenius14}. While the beaming parameter has a dependence on the phase angle \citep[see, e.g.,][]{Delbo2003,AliLagoa2018}, this cannot be considered in our cases due to the very limited phase angle ranges.

\subsection{Colors}

In the final stage of our investigation we also used color information of our targets. Following the scheme in \citet{Lacerda2014} we use the spectral slope S$^\prime$ in units of [\%/(1000\,$\AA$)], to quantify the visible colors. This spectral slope is calculated as 
\citep{Hainaut2002}

\begin{equation}
S^\prime = 100\frac{R(\lambda_2)-R(\lambda_1)}{(\lambda_2-\lambda_1)/1000}
\label{eq:slope01}
\end{equation}
\begin{equation}
R(\lambda) = 10^{-0.4[ (m(\lambda)-m_{\odot}(\lambda)) -  (m(V)-m_{\odot}(V)) ]}
\label{eq:slope02}
,\end{equation}

\noindent where $R$ is the reflectance normalized to the $V$ band, and  $m$ and $m_{\odot}$ are the magnitudes of the object and of the Sun at the wavelength $\lambda$.

Broadband colors are taken from the literature; these observations were usually executed in the Johnson-Cousins B, V, R, I bands or the SLOAN g$^\prime$, r$^\prime$, i$^\prime$ bands. In Table~\ref{table:colours} we present the V-R colors, the S$^\prime$ spectral slopes, and the source of the data. SLOAN colors were transformed to Johnson-Cousins colors following \citet{Sheppard2012}. 

\begin{table}[ht!]
\caption{Absolute brightness values and colors of our targets. The columns are (from left to right)  V-band absolute magnitude of the target; $V-R$ color index; reference of \absmag \, and $V-R$; spectral slope.
{\it References:} (1) \cite{Dor2007}; (2) \cite{Dor2005}; (3) \cite{SS09}; (4) \cite{Sheppard2012}; (5) \cite{Perna2010}; (6) \cite{Sheppard02}; (7) Derived from MPC V-data; (8) \cite{Brucker2009}; (9) \cite{Boehnhardt}; (10) \cite{Barucci99}. Objects flagged with  an asterisk $^*$  have a known light curve amplitude which was considered as an additional source of \absmag\, uncertainty \citep[see also][and Sect~\ref{sect:individual}]{Vilenius14}.}
\centering
\scriptsize
\begin{tabular}{c|cccc}
\hline
object & $H_V$ & V-R & Ref. & S$^\prime$  \\
       & (mag) & (mag) &    & $\% /(1000\,\AA)$ \\
\hline
1995\,QY$_9$ & 7.49$\pm$0.30 & 0.47$\pm$0.12 & 10 & 11.60$\pm$12.50\\[1ex]
1996\,GQ$_{21}$ & 5.50$\pm$0.05 & 0.73$\pm$0.04 & 1 & 38.70$\pm$4.10\\[1ex]
2000\,CN$_{105}$ & 5.68$\pm$0.06 & 0.66$\pm$0.08 & 3 & 31.48$\pm$8.25\\[1ex]
2000\,YW$_{134}$ & 4.65 $\pm$0.06 & 0.45$\pm$0.08 & 3 & 9.50$\pm$8.40\\[1ex]
2001\,FZ$_{173}$ & 6.30$\pm$0.01 & 0.58$\pm$0.04 & 1 & 23.15$\pm$4.15\\[1ex]
2001\,QG$_{298}$$^*$ & 6.81$\pm$0.03 & 0.70$\pm$0.04 & 9 & 35.62$\pm$4.13\\[1ex]
2001\,QR$_{322}$ & 8.11$\pm$0.02 & 0.46$\pm$0.02 & 8 & 10.55$\pm$2.11 \\[1ex]
2001\,QX$_{322}$ & 6.55$\pm$0.10 & 0.65$\pm$0.11 & 3 & 30.45$\pm$11.3\\[1ex]
2001\,UR$_{163}$ & 4.46$\pm$0.02 & 0.84$\pm$0.03 & 1 & 49.88$\pm$3.01\\[1ex]
2001\,YH$_{140}$ & 5.72$\pm$0.11 & 0.56$\pm$0.02 & 4 & 21.07$\pm$2.09\\[1ex]
2002\,CY$_{224}$ & 6.35$\pm$0.05 & 0.66$\pm$0.06 & 3 & 31.48$\pm$6.10\\[1ex]
2002\,GP$_{32}$ & 6.9$\pm$0.11 & 0.34$\pm$0.06 & 2 & 11.61$\pm$4.21\\[1ex]
2003\,FB$_{128}$ & 7.26$\pm$0.05 & 0.50$\pm$0.06 & 9 & 14.77$\pm$6.32\\[1ex]
2003\,FE$_{128}$ & 6.94$\pm$0.07 & 0.68$\pm$0.08 & 9 & 33.56$\pm$8.30\\[1ex]
2003\,QX$_{111}$ & 6.76$\pm$0.50 & -- & 8 & -- \\[1ex]
2003\,UY$_{117}$ & 5.91$\pm$0.04 & 0.56$\pm$0.01 & 4 & 21.07$\pm$1.10\\[1ex]
2003\,UZ$_{413}$$^*$ & 4.38$\pm$0.05 & 0.45$\pm$0.04 & 5 & 9.50$\pm$4.70\\[1ex]
2003\,WU$_{172}$ & 6.70$\pm$0.37 & -- & 7 & --\\[1ex]
2004\,PG$_{115}$ & 5.46$\pm$0.05 & 0.31$\pm$0.08 & 9 & -5.28$\pm$8.43\\[1ex]
2005\,RM$_{43}$$^*$ & 4.52$\pm$0.01 & 0.33$\pm$0.02 & 9 & -3.17$\pm$2.11\\[1ex]
2005\,RS$_{43}$ & 5.14$\pm$0.03 & 0.46$\pm$0.03 & 9 & 10.55$\pm$3.16\\[1ex]
\hline
1999\,KR$_{16}$ & 5.37$\pm$0.02 & 0.75$\pm$0.04 & 6 & 40.75$\pm$4.10\\[1ex]
2005\,QU$_{182}$ & 3.99$\pm$0.02 & 0.54$\pm$0.03 & 9 & 18.97$\pm$3.15\\[1ex]
\hline
\end{tabular}
\label{table:colours}
\end{table}


\section{Radiometric model results \label{sect:results}}

\subsection{General results \label{sect:generalresults}}

We derived new albedo, effective diameter, and beaming parameter values for 20 trans-Neptunian resonant and scattered disk objects via NEATM radiometric modeling using mainly Herschel/PACS data, in some cases supplemented by Spitzer/MIPS measurements, and for 3 targets based solely on Spitzer/MIPS data. The results are presented in Table~\ref{table:results}, and the best-fit radiometric model fits are shown in Figs.~\ref{fig:seds}~and~\ref{fig:seds2}. Due to non-detections (upper limits) of the thermal emission at all bands in the case of 2002\,CY$_{224}$ we were only able to derive an upper limit for the effective diameter and a lower limit for the geometric albedo. 

\begin{figure}
    \centering
    \includegraphics[width=\columnwidth]{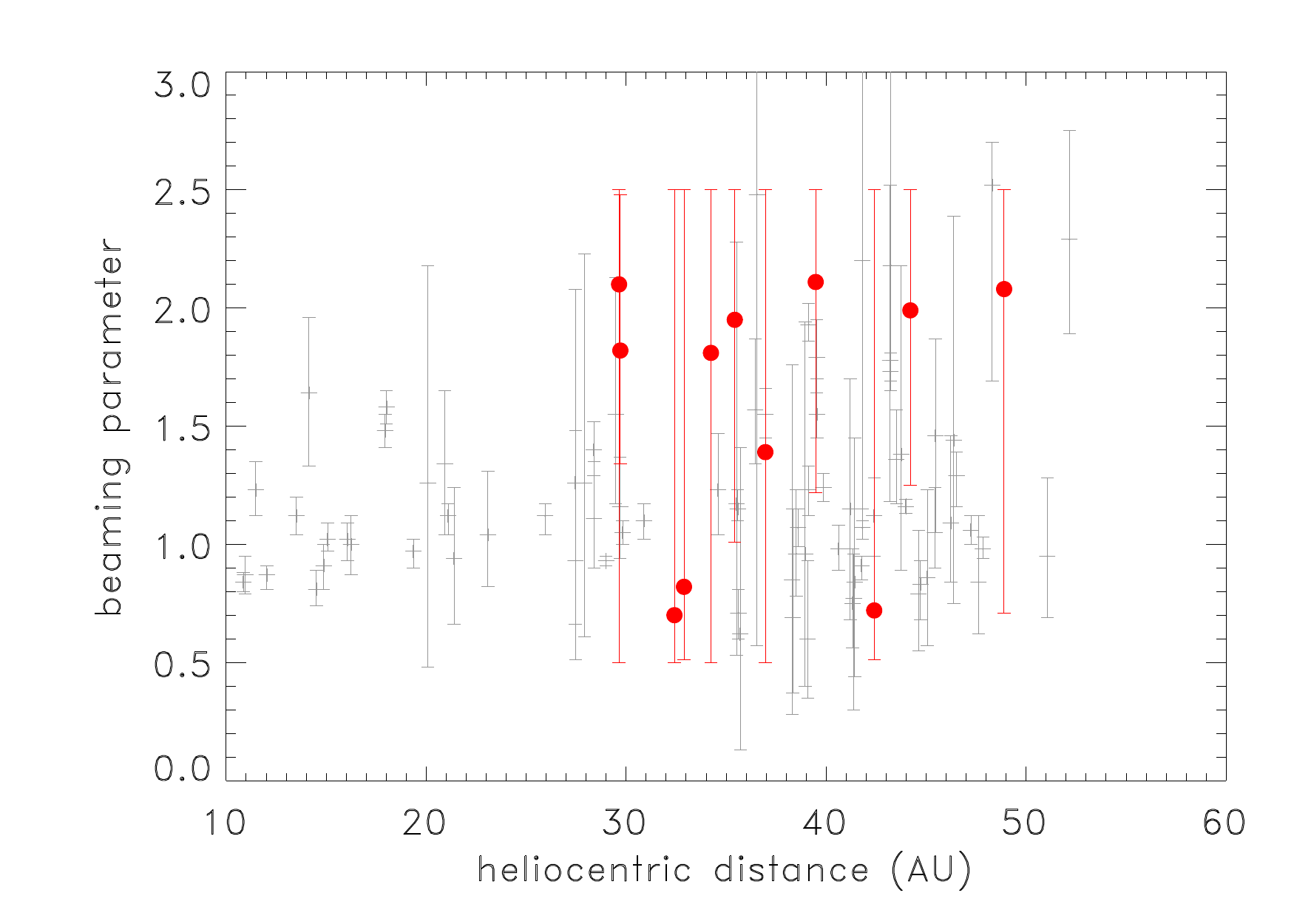}
    \caption{Beaming parameter vs. the heliocentric distance of the target at the time of the observation. Red symbols indicate the targets from the present paper for which the beaming parameter could be derived within the predefined limits. Gray symbols in the background indicate the beaming parameters from \citet{Lellouch2013}. }
    \label{fig:eta_vs_rh}
\end{figure}

The fitted beaming parameters  of those targets for which we did {not} use fixed-$\eta$ are plotted in  Fig.~\ref{fig:eta_vs_rh} as a function of the heliocentric distance (red symbols). Compared with the \citet{Lellouch2013} beaming parameters (gray symbols in Fig.~\ref{fig:eta_vs_rh}), in general  the new targets present  higher $\eta$ values, with a mean value of $\eta$\,=\,1.56 and  standard deviation of d$\eta$\,=\,0.53, while this is 1.18 for the \citet{Lellouch2013} objects. A sample compiled from the \citet{Lellouch2013} targets with heliocentric distances of r$_h$\,$\geq$\,29\,AU and from our floating-$\eta$ fit targets gives an average beaming parameter of $\eta$\,=\,1.25, and we used this value for the fixed-$\eta$ radiometric fitting (see Sect.~\ref{sect:radmod}). 

It should be noted that objects presented in this paper are among the faintest ones in the ``TNOs are Cool'' sample. This is reflected in the large flux density uncertainties, and also in the radiometric modeling in the large $\eta$ uncertainties for many objects;  in these cases the observed fluxes could be fitted with a large range of beaming parameters, as  also shown in Figs.~\ref{fig:seds}~and~\ref{fig:seds2}. 

\begin{table}[ht!]
\caption{Radiometric fit results of our new targets. Column headings are  (from left to right) object name; Data: P:PACS, M:MIPS; Diam: diameter of object; p$_V$: geometric albedo; $\eta$: beaming parameter; 
*: fixed $\eta$ was used. 
}
\label{table:results}
\centering
\scriptsize
\begin{tabular}{c|cccccc}
\hline
Object & Data & Diam (km) & Pv & $\eta$   \\
\hline
\hline
\\
(32929) 1995\,QY$_{9}$   & M     & $< 210$              & $>0.022$ & $1.25^*\pm 0.35$    \\[1ex]
(26181) 1996\,GQ$_{21}$  & P/M   & $456^{+89}_{-105}$   & $0.060^{+0.040}_{-0.022}$ & $2.41^{+0.09}_{-1.40}$  \\[1ex]
                                                & P/M    & $352^{+46}_{-61}$    & $0.098^{+0.046}_{-0.030}$ & $1.25^*\pm 0.35$ \\[1ex]
(40314) 1999\,KR$_{16}$  & P     & $202^{+108}_{-36}$   & $0.307^{+0.156}_{-0.178}$ & $0.53^{+1.84}_{-0.02}$ \\[1ex]
                                                 & P     & $271^{+39}_{-45}$    & $0.177^{+0.076}_{-0.048}$ & $1.25^*\pm 0.35$ \\[1ex]
(523588) 2000\,CN$_{105}$ & P    & $< 440$ & $> 0.044$ & $1.25^*\pm 0.35$  \\[1ex]
(82075) 2000\,YW$_{134}$ & P/M   & $437^{+118}_{-137}$  & $0.133^{+0.173}_{-0.051}$ & $1.99^{+0.51}_{-0.74}$  \\[1ex]
(82155) 2001\,FZ$_{173}$     & P     & $212^{+157}_{42}$    & $0.121^{+0.065}_{-0.082}$ & $0.70^{+1.80}_{-0.20}$    \\[1ex]
(139775) 2001\,QG$_{298}$ & P    & $< 215$      & $> 0.070$ & $1.25^*\pm 0.35$   \\[1ex]
         2001\,QR$_{322}$ & M    & $178^{+61}_{-37}$    & $0.033^{+0.021}_{-0.015}$ & $1.82^{+0.66}_{-0.48}$  \\[1ex]
        2001\,QX$_{322}$  & P/M  & $< 210$      & $>0.080$ & $1.25^*\pm 0.35$        \\[1ex]
(42301) 2001\,UR$_{163}$  &  P/M & $261^{+209}_{-91}$ & $0.427^{+0.624}_{-0.297}$ & $0.59^{+1.9}_{-0.09}$ \\[1ex]
                                                & P/M & $367^{+103}_{-197}$ & $0.220^{+0.832}_{-0.090}$ & $1.25^*\pm 0.35$ \\[1ex]
(126154) 2001\,YH$_{140}$ & P/M  & $252^{+148}_{-52}$   & $0.143^{+0.144}_{-0.094}$ & $1.24^{+1.26}_{-0.62}$   \\[1ex]
(119878) 2002\,CY$_{224}$ & P    & $< 220$      & $> 0.098$ & $1.25^*\pm 0.35$   \\[1ex]
        2002\,GP$_{32}$  & P     & $234^{+47}_{-84}$    & $0.063^{+0.105}_{-0.026}$ & $2.47^{+0.03}_{-1.90}$   \\[1ex]
                                        & P      & $181^{+37}_{-47}$    & $0.096^{+0.105}_{-0.040}$ & $1.25^*\pm 0.35$   \\[1ex] 
(133067) 2003\,FB$_{128}$ & P    & $218^{+91}_{-108}$   & $0.047^{+0.144}_{-0.025}$ & $1.81^{+0.69}_{-1.31}$  \\[1ex]
(469505) 2003\,FE$_{128}$ & P    & $137^{+123}_{-37}$   & $0.184^{+0.160}_{-0.145}$ & $0.51^{+1.99}_{-0.01}$  \\[1ex]
                                            & P          & $189^{+71}_{-89}$    & $0.079^{+0.266}_{-0.040}$ & $1.25^*\pm 0.35$    \\[1ex]
                2003\,QX$_{111}$ & M     & $293^{106}_{-93}$    & $0.033^{+0.164}_{-0.025}$ & $2.11^{+0.39}_{-0.89}$   \\[1ex]
(143707) 2003\,UY$_{117}$ & P    & $196^{+114}_{-54}$    & $0.182^{+0.195}_{-0.103}$ & $0.82^{+1.68}_{-0.31}$ \\[1ex]
(455502) 2003\,UZ$_{413}$ & P    & $472^{+122}_{-25}$   & $0.151^{+0.025}_{-0.064}$ & $0.53^{+0.42}_{-0.03}$ \\ [1ex]
                                                        & P      & $650^{+1}_{-175}$    & $0.075^{+0.076}_{-0.006}$ & $1.25^*\pm 0.35$ \\ [1ex]
(450265) 2003\,WU$_{172}$ & P    & $261^{+48}_{-114}$   & $0.074^{+0.171}_{-0.050}$ & $2.10^{+0.40}_{-1.60}$   \\[1ex]
(307982) 2004\,PG$_{115}$ & P    & $334^{+1}_{-191}$    & $0.101^{+0.534}_{-0.005}$ & $1.39^{+1.11}_{-0.89}$   \\[1ex]
(303775) 2005\,QU$_{182}$ & P    & $584^{+155}_{-144}$  & $0.129^{+0.115}_{-0.046}$ & $2.08^{+0.42}_{-1.37}$          \\[1ex]
(145451) 2005\,RM$_{43}$ & P     & $524^{+96}_{-103}$   & $0.102^{+0.057}_{-0.029}$ & $1.95^{+0.55}_{-0.94}$   \\[1ex]
(308379) 2005\,RS$_{43}$ & P     & $228^{+112}_{-58}$   & $0.311^{+0.247}_{-0.182}$ & $0.72^{+1.78}_{-0.21}$ \\[1ex]
\hline
\end{tabular}
\end{table}

\begin{figure*}[!ht]
\hbox{
\resizebox{5.5cm}{!}{\rotatebox{0}{\includegraphics{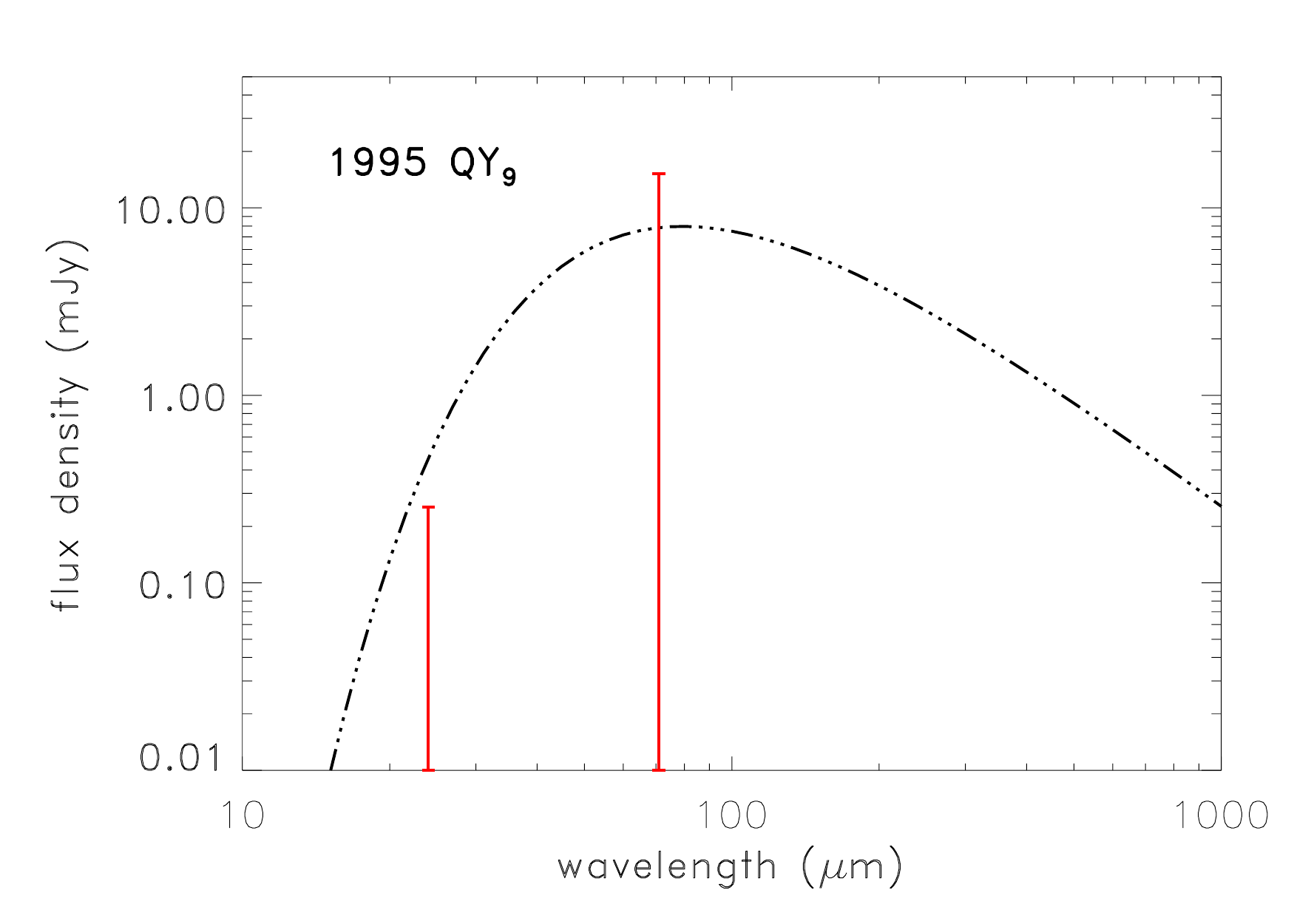}}}
\resizebox{5.5cm}{!}{\rotatebox{0}{\includegraphics{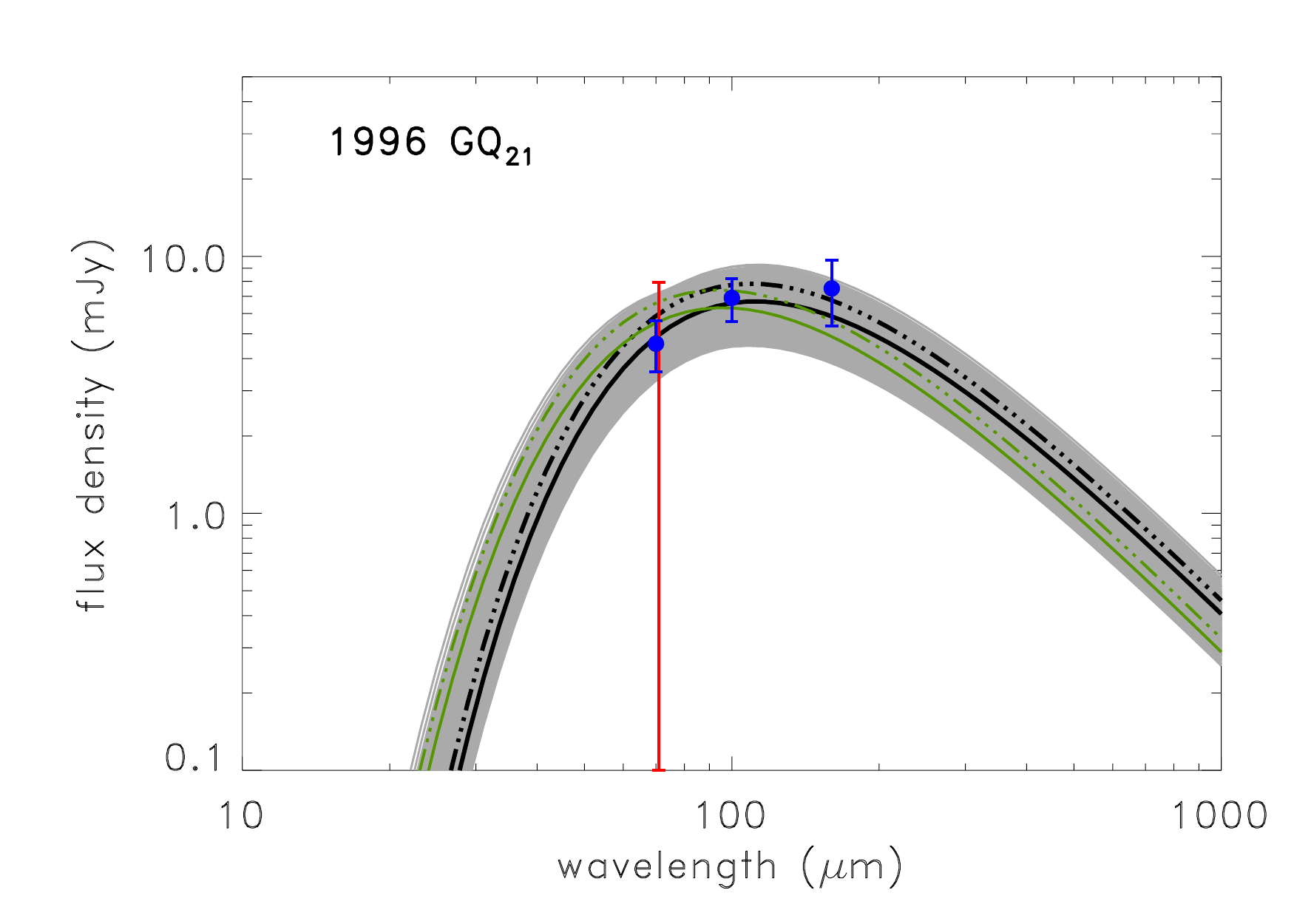}}}
\resizebox{5.5cm}{!}{\rotatebox{0}{\includegraphics{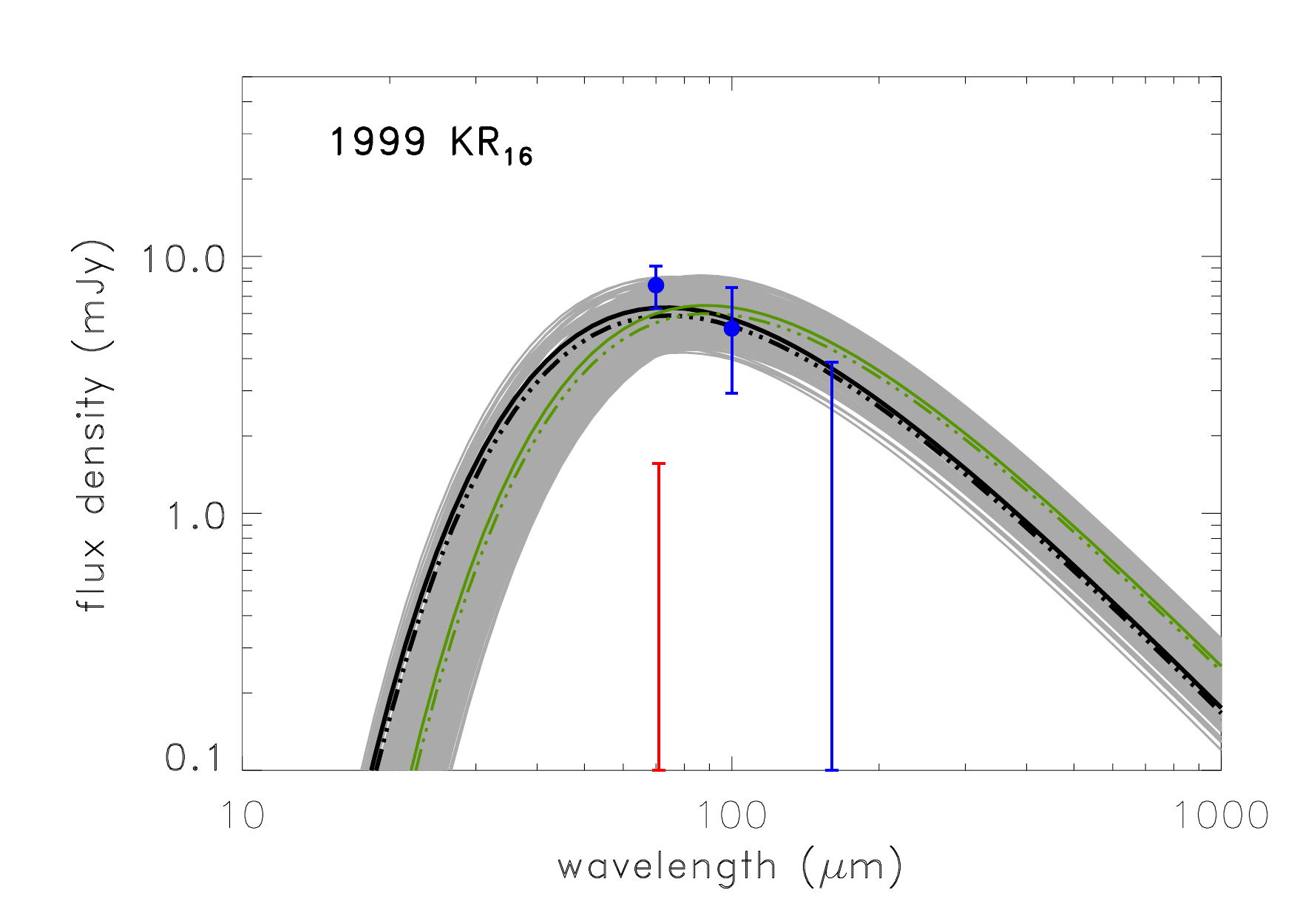}}}
}
\hbox{
\resizebox{5.5cm}{!}{\rotatebox{0}{\includegraphics{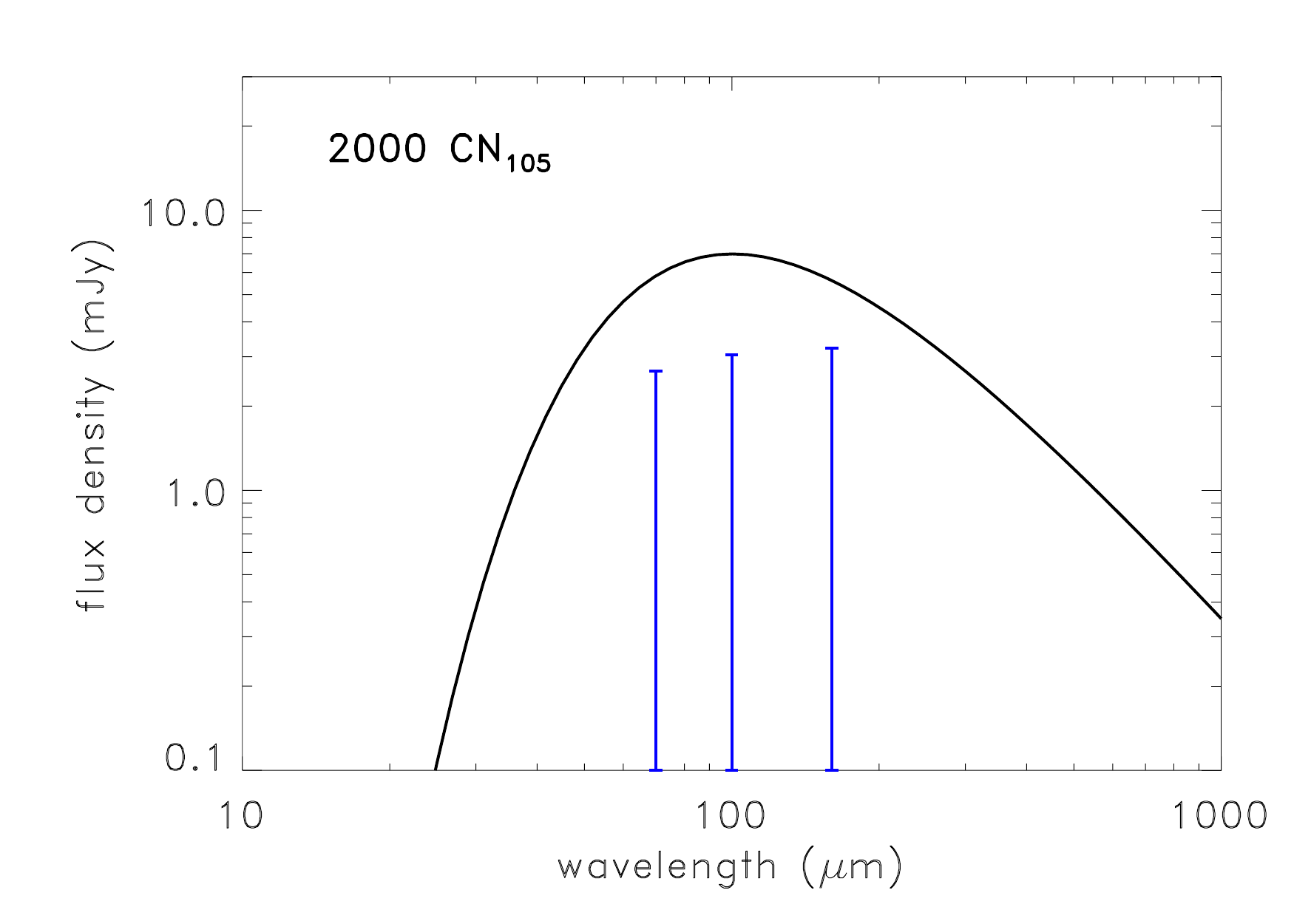}}}
\resizebox{5.5cm}{!}{\rotatebox{0}{\includegraphics{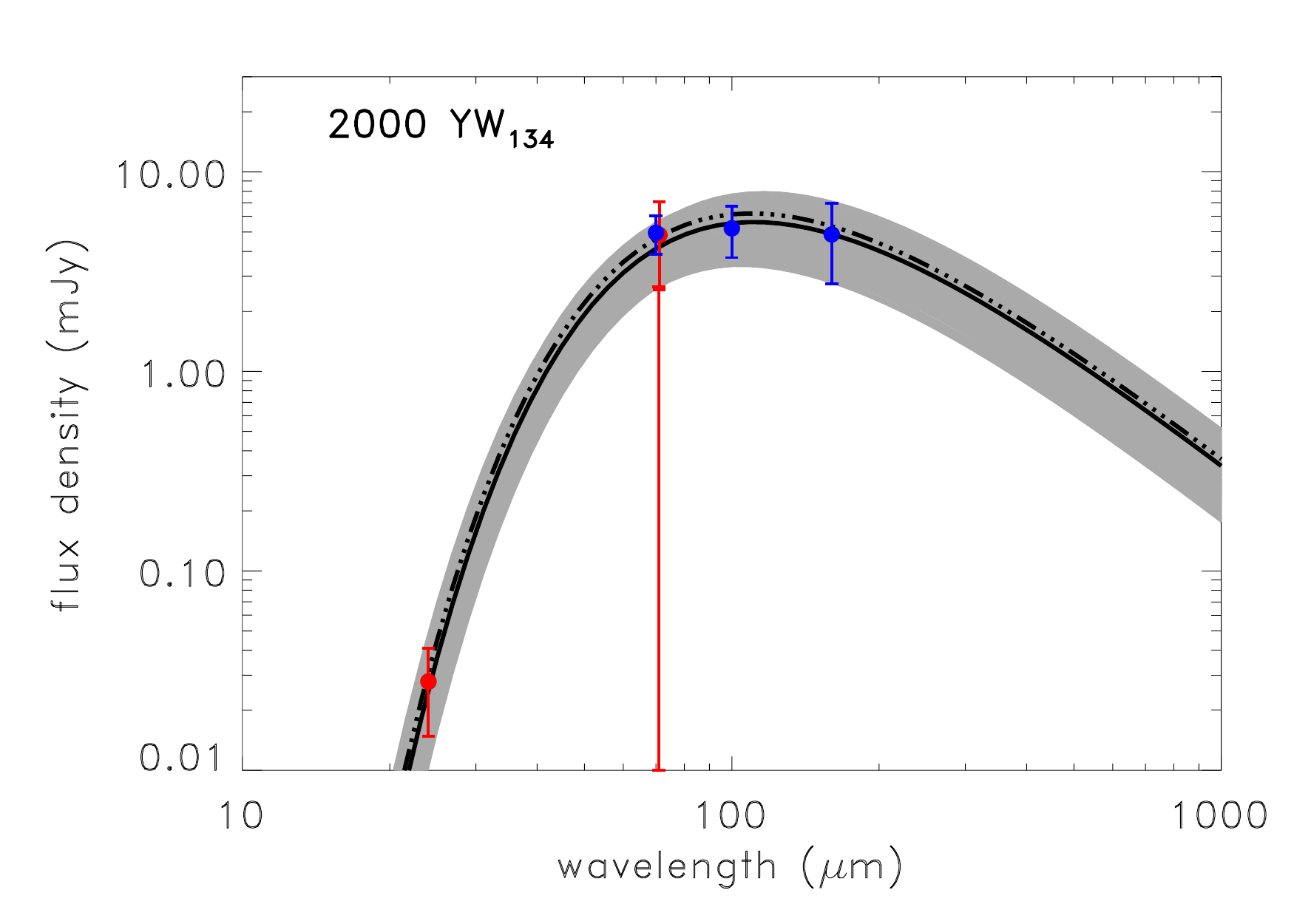}}}
\resizebox{5.5cm}{!}{\rotatebox{0}{\includegraphics{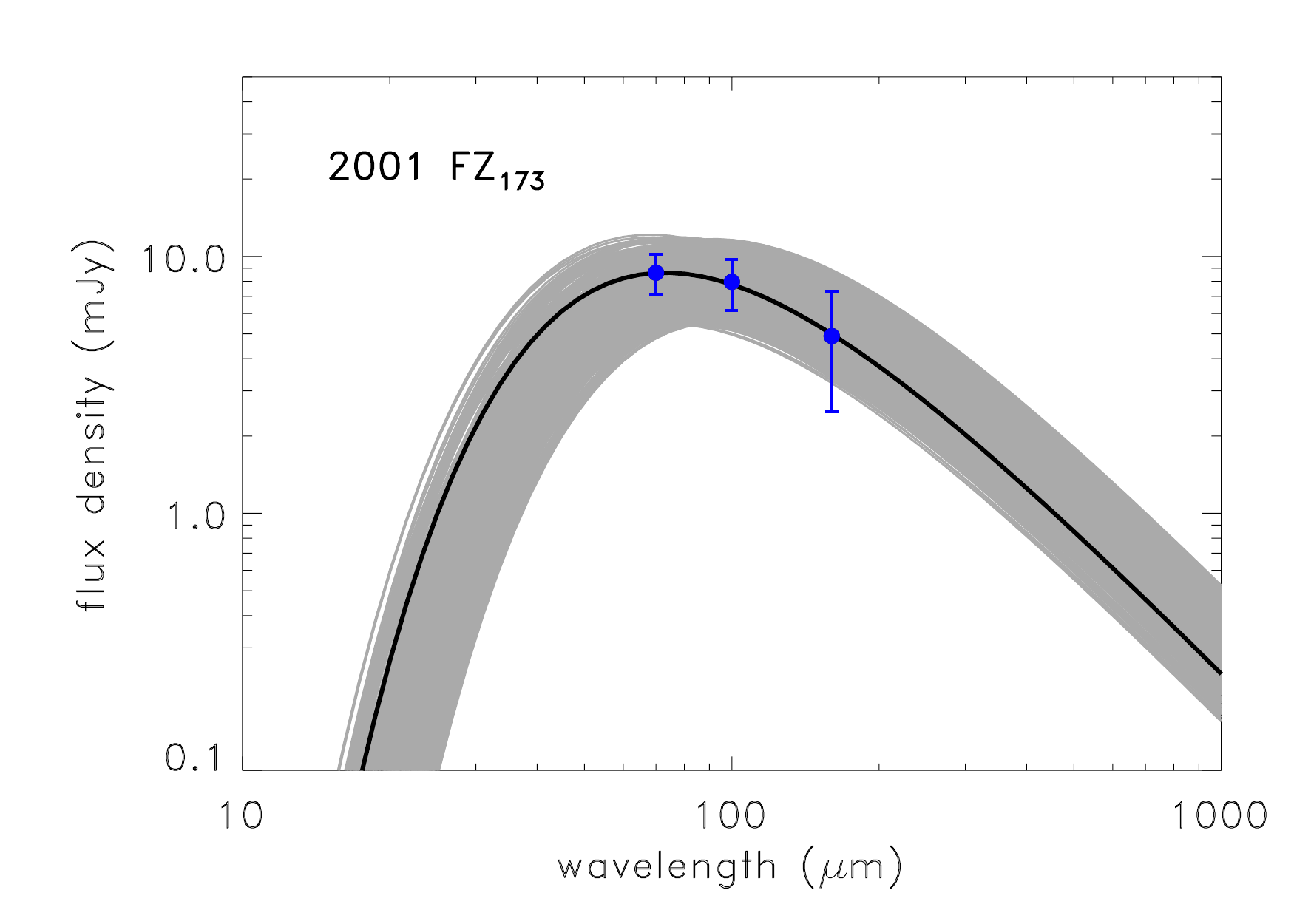}}}
}
\hbox{
\resizebox{5.5cm}{!}{\rotatebox{0}{\includegraphics{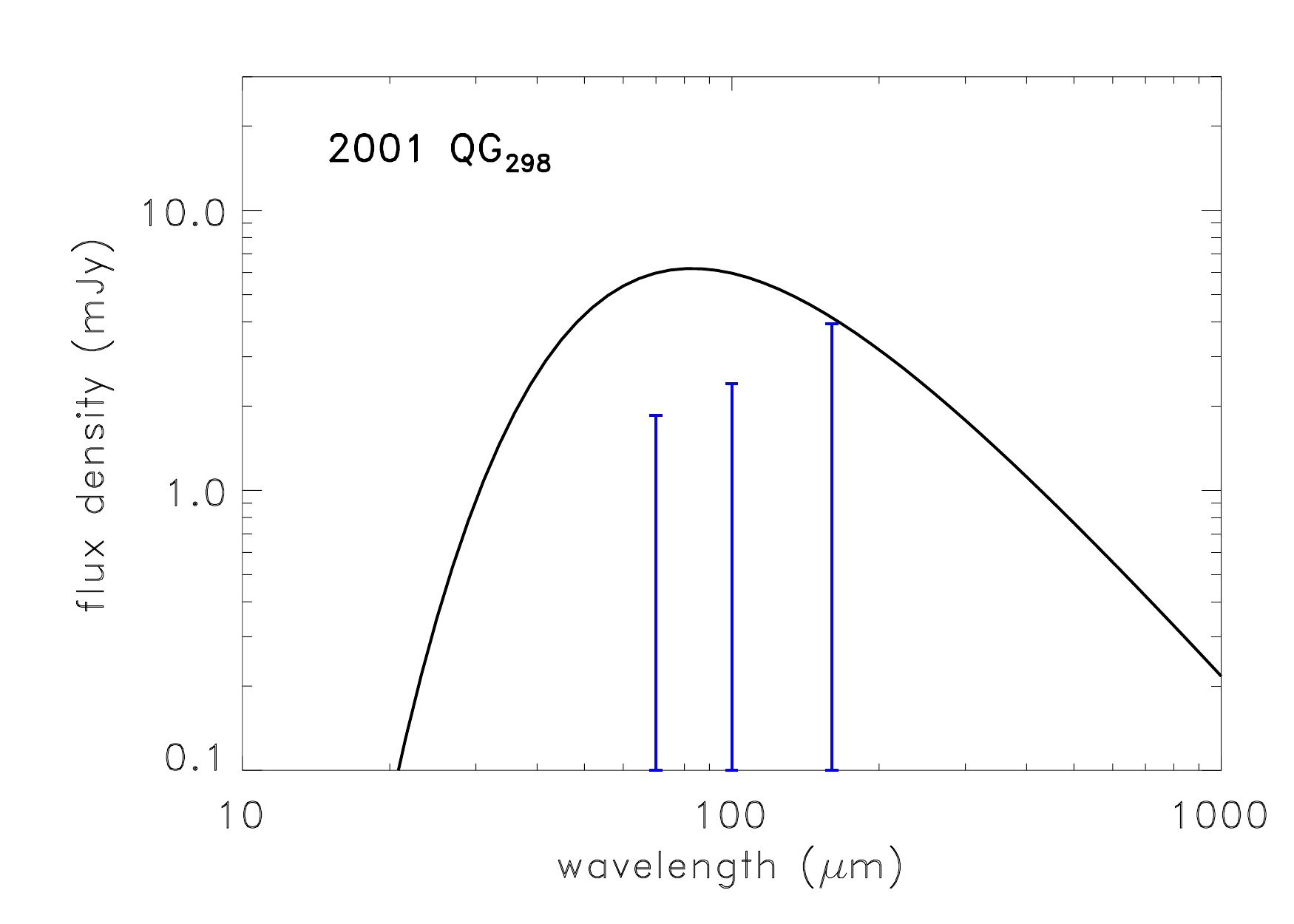}}}
\resizebox{5.5cm}{!}{\rotatebox{0}{\includegraphics{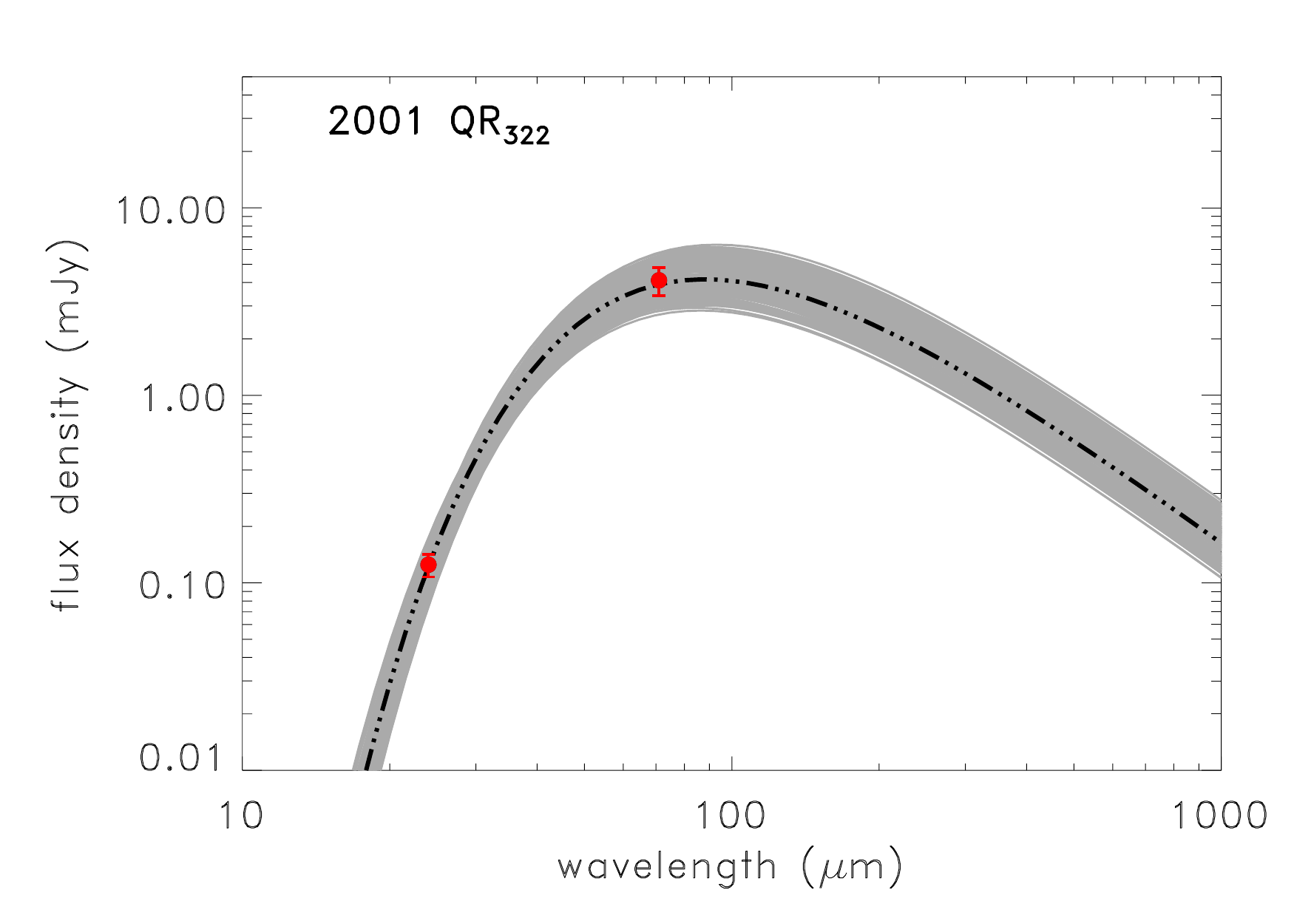}}}
\resizebox{5.5cm}{!}{\rotatebox{0}{\includegraphics{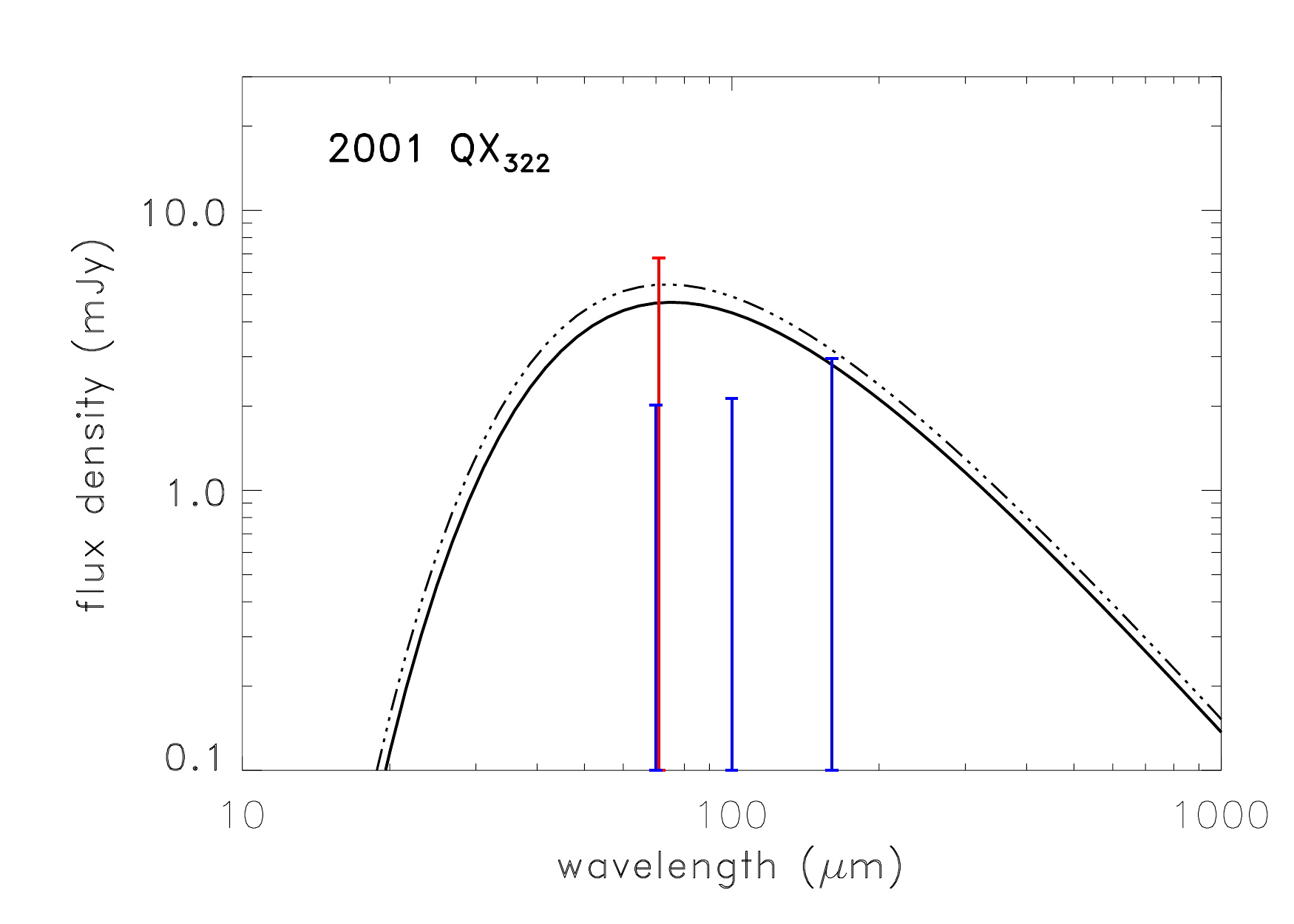}}}
}
\hbox{
\resizebox{5.5cm}{!}{\rotatebox{0}{\includegraphics{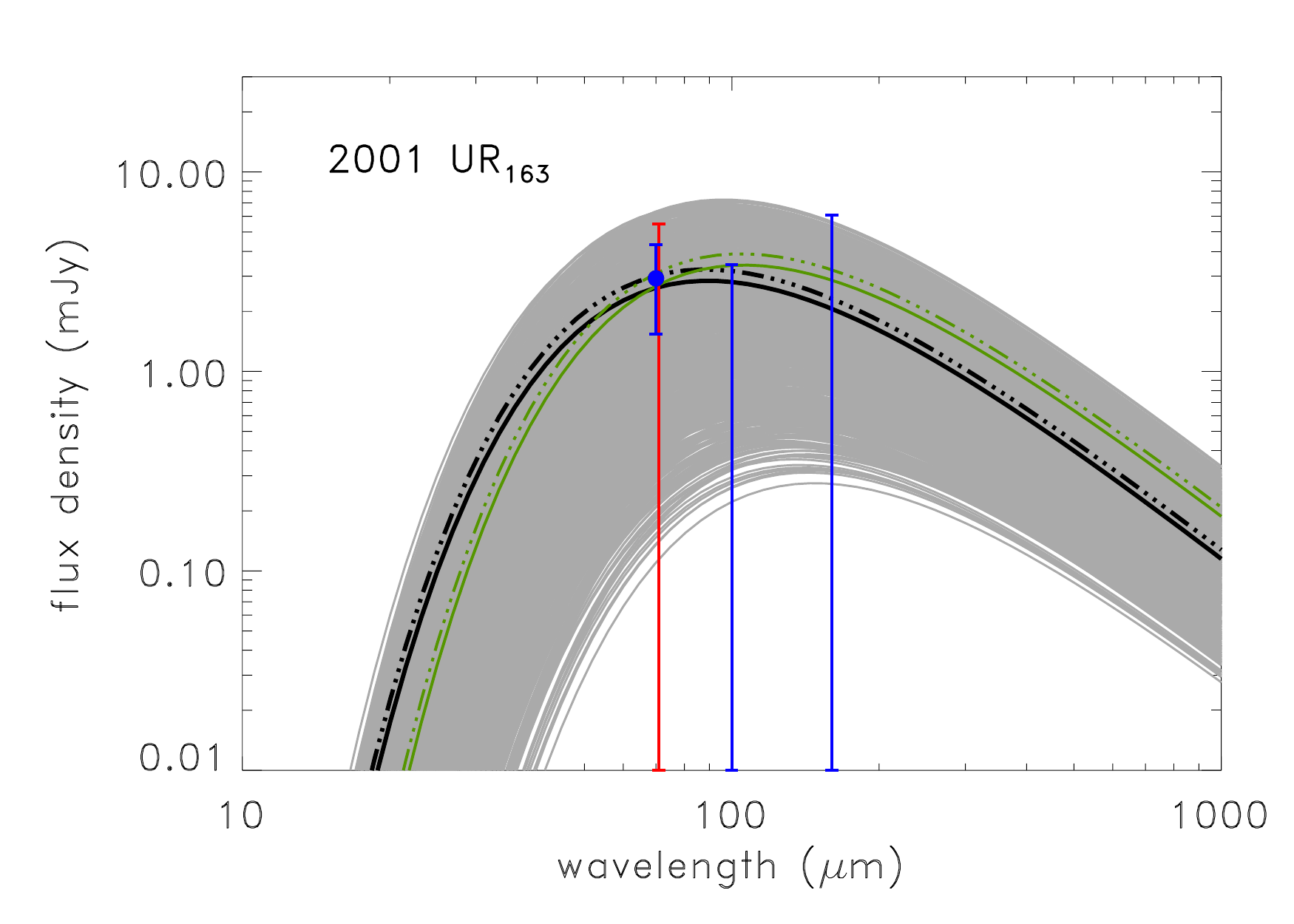}}}
\resizebox{5.5cm}{!}{\rotatebox{0}{\includegraphics{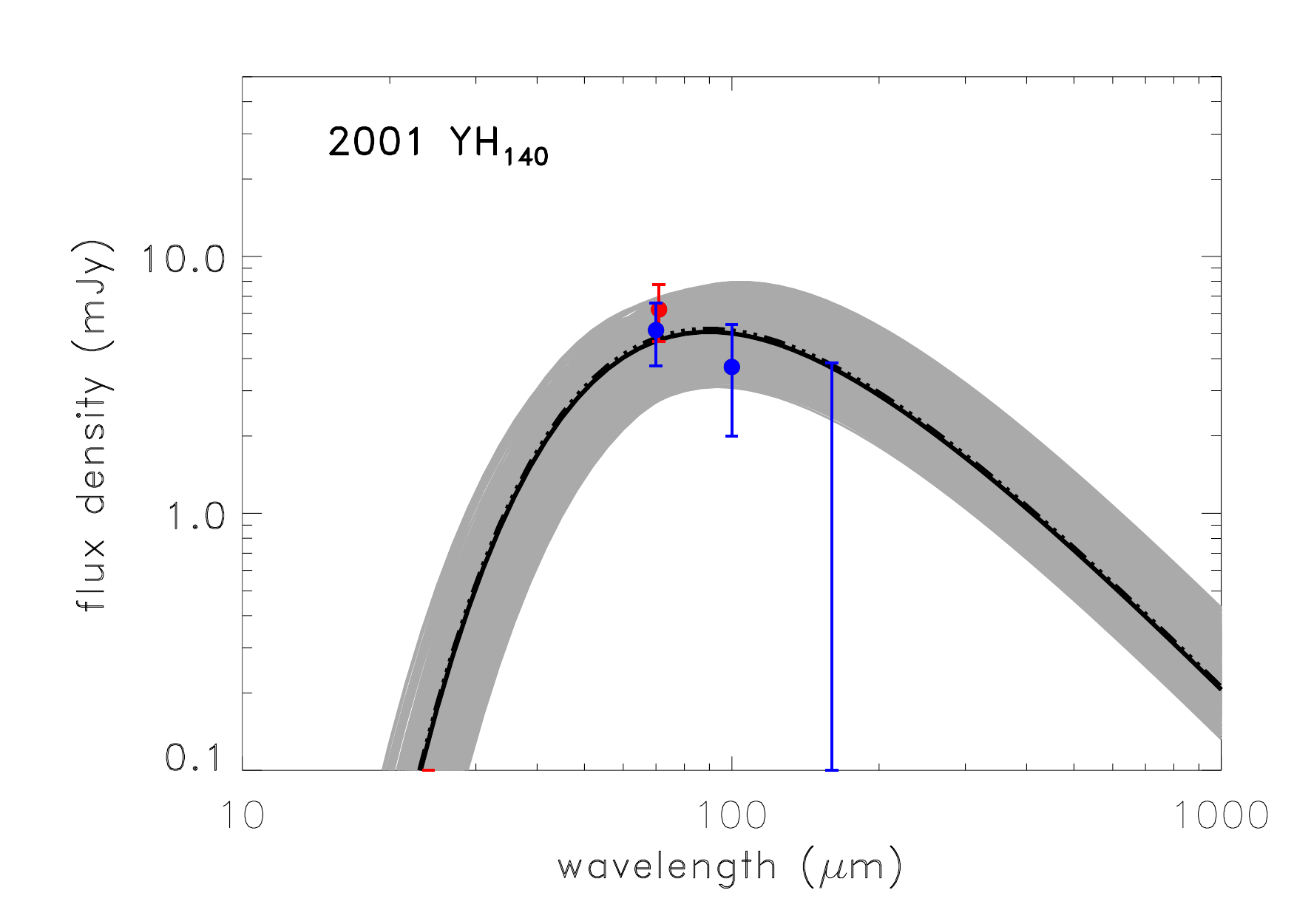}}}
\resizebox{5.5cm}{!}{\rotatebox{0}{\includegraphics{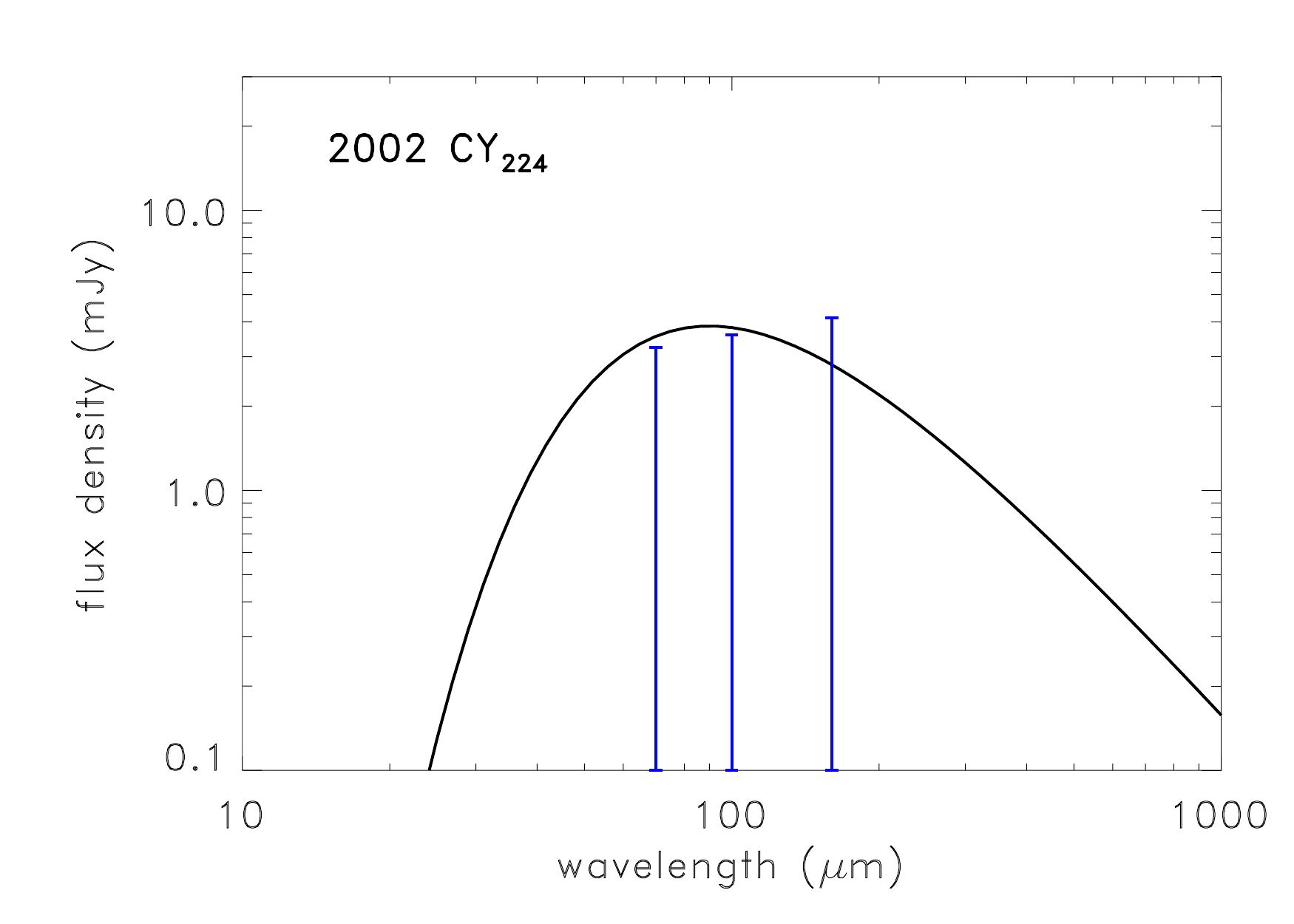}}}
}
\caption{NEATM fits of the thermal emission of our target sample. Black solid and dash-dotted curves represent the spectral energy distribution of the best-fit (lowest $\chi^2$) model for the epochs of the PACS and MIPS measurements, respectively. The gray area shows the zone of NEATM model curves that are compatible with the observed measurement uncertainties. In those cases when the floating $\eta$ fits converged to one of the beaming parameter limits, the fit was repeated with a fixed beaming parameter of $\eta$\,=\,1.25, represented by the green curves (not in all panels). Red and blue symbols represent the measured MIPS and PACS flux densities, respectively, color-corrected according to the best-fit NEATM model, at their respective epochs. In those cases when only flux density upper limits were available, we show the SED of the NEATM model with the diameter and geometric albedo limits presented in Table~\ref{table:results}.} \label{fig:seds}
\end{figure*}
\begin{figure*}[ht!]
\hbox{
\resizebox{5.5cm}{!}{\rotatebox{0}{\includegraphics{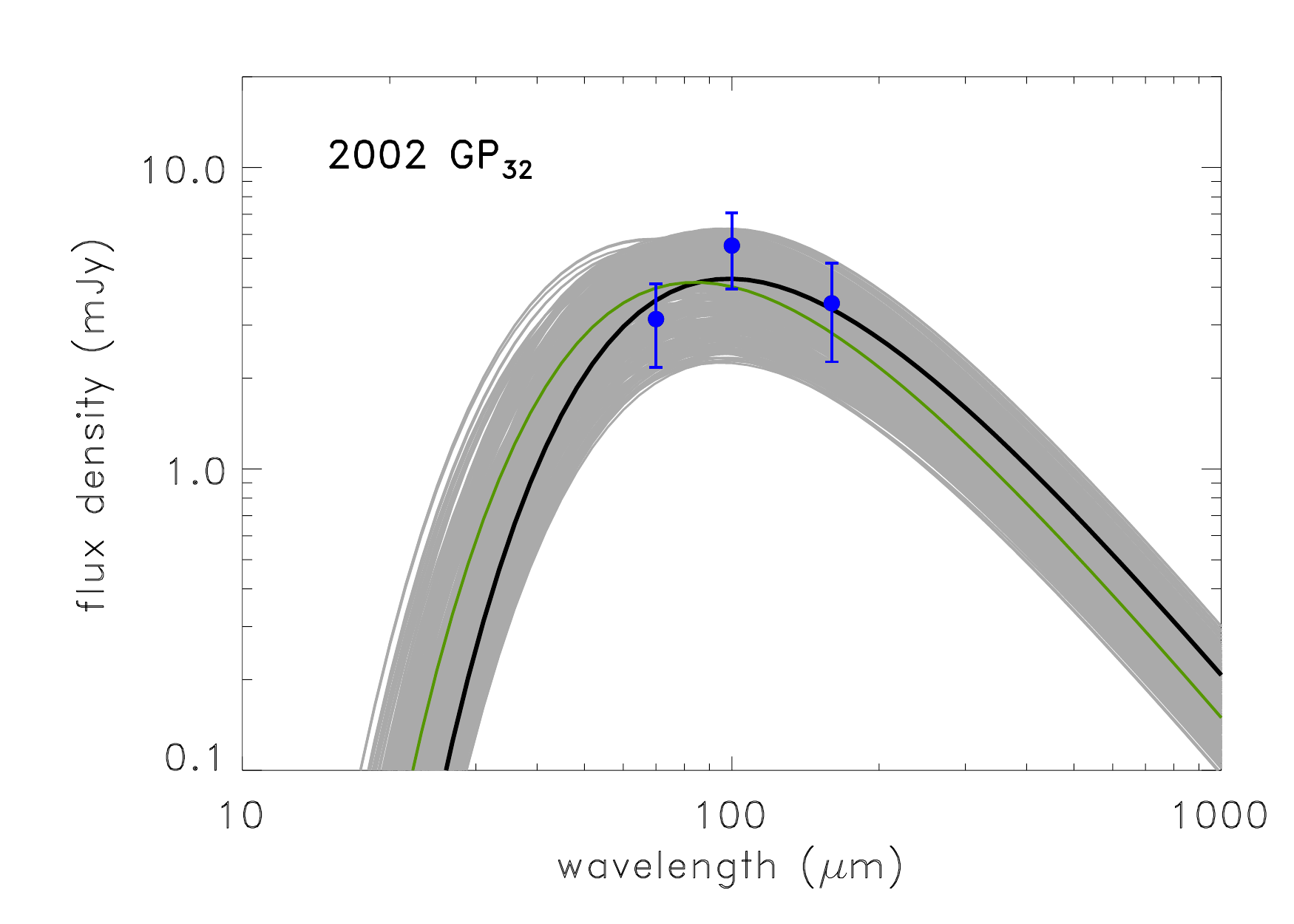}}}
\resizebox{5.5cm}{!}{\rotatebox{0}{\includegraphics{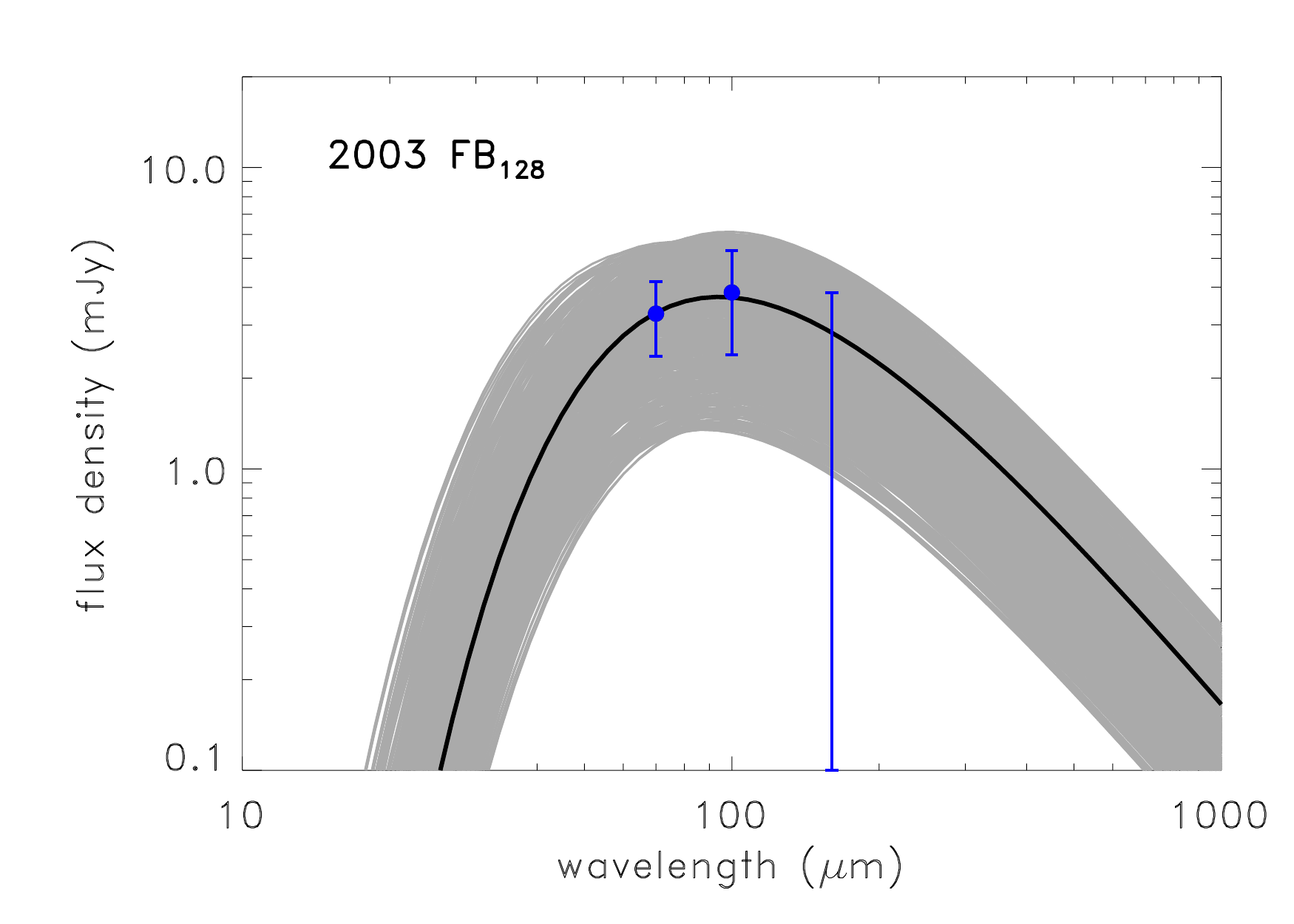}}}
\resizebox{5.5cm}{!}{\rotatebox{0}{\includegraphics{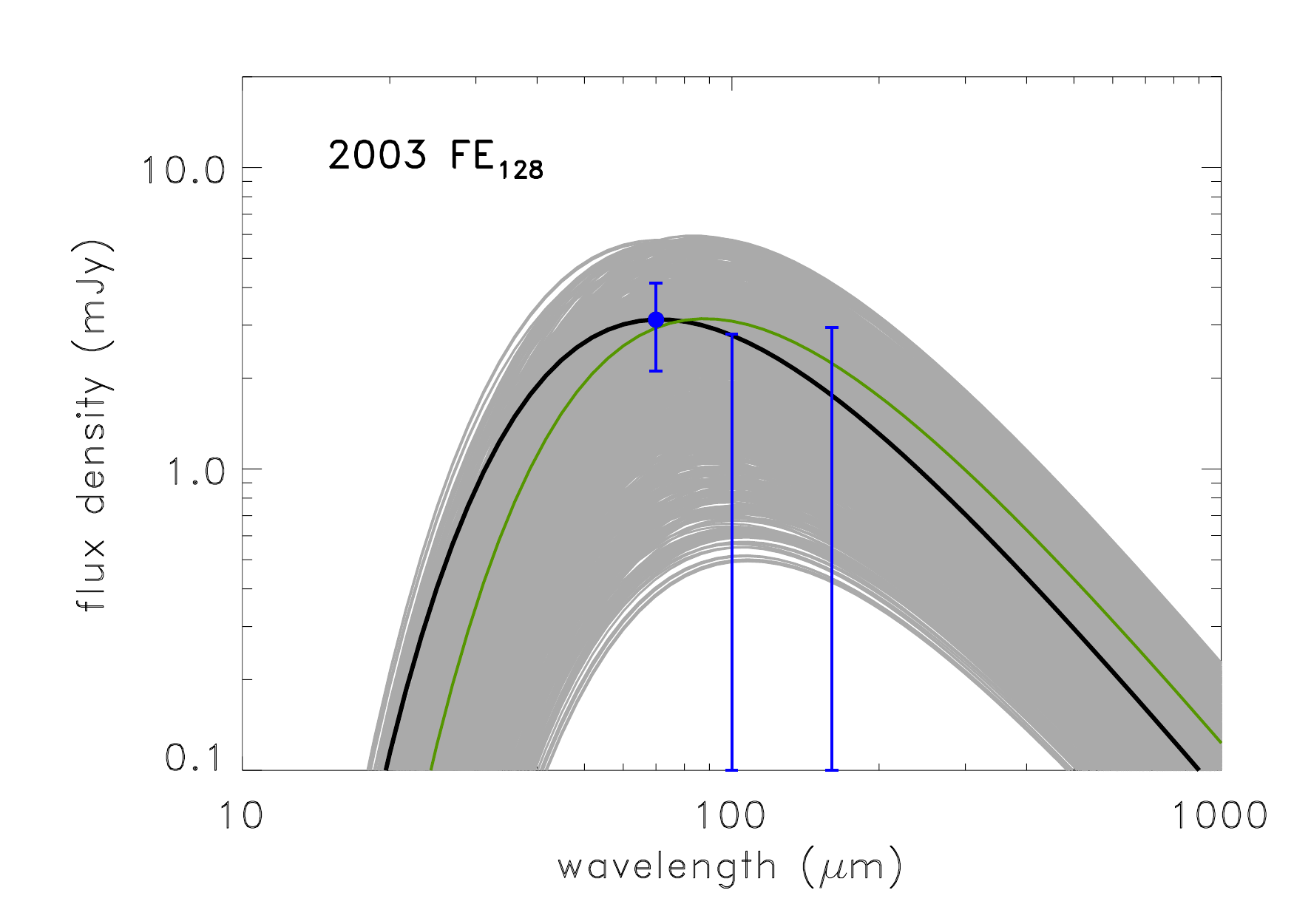}}}
}
\hbox{
\resizebox{5.5cm}{!}{\rotatebox{0}{\includegraphics{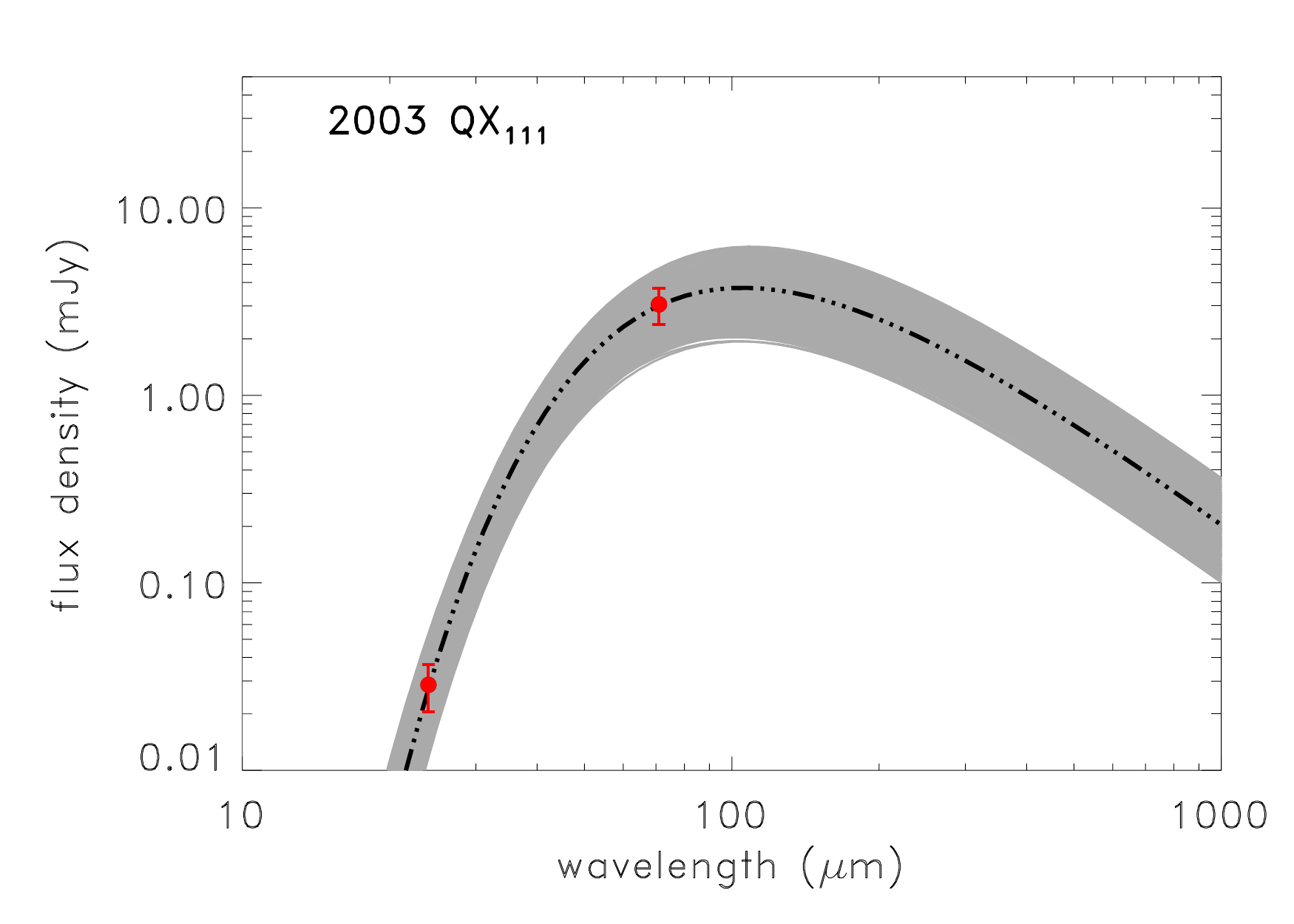}}}
\resizebox{5.5cm}{!}{\rotatebox{0}{\includegraphics{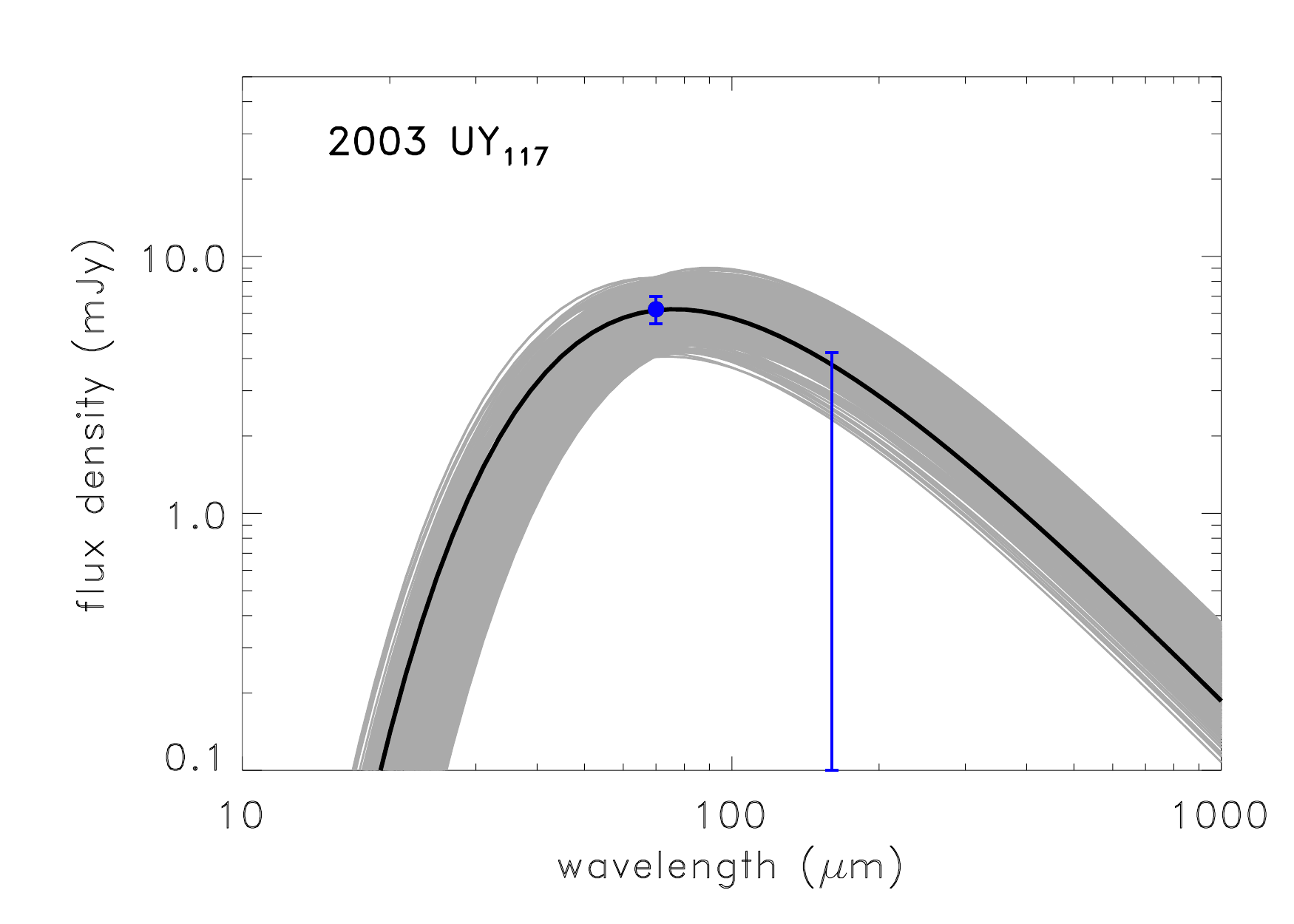}}}
\resizebox{5.5cm}{!}{\rotatebox{0}{\includegraphics{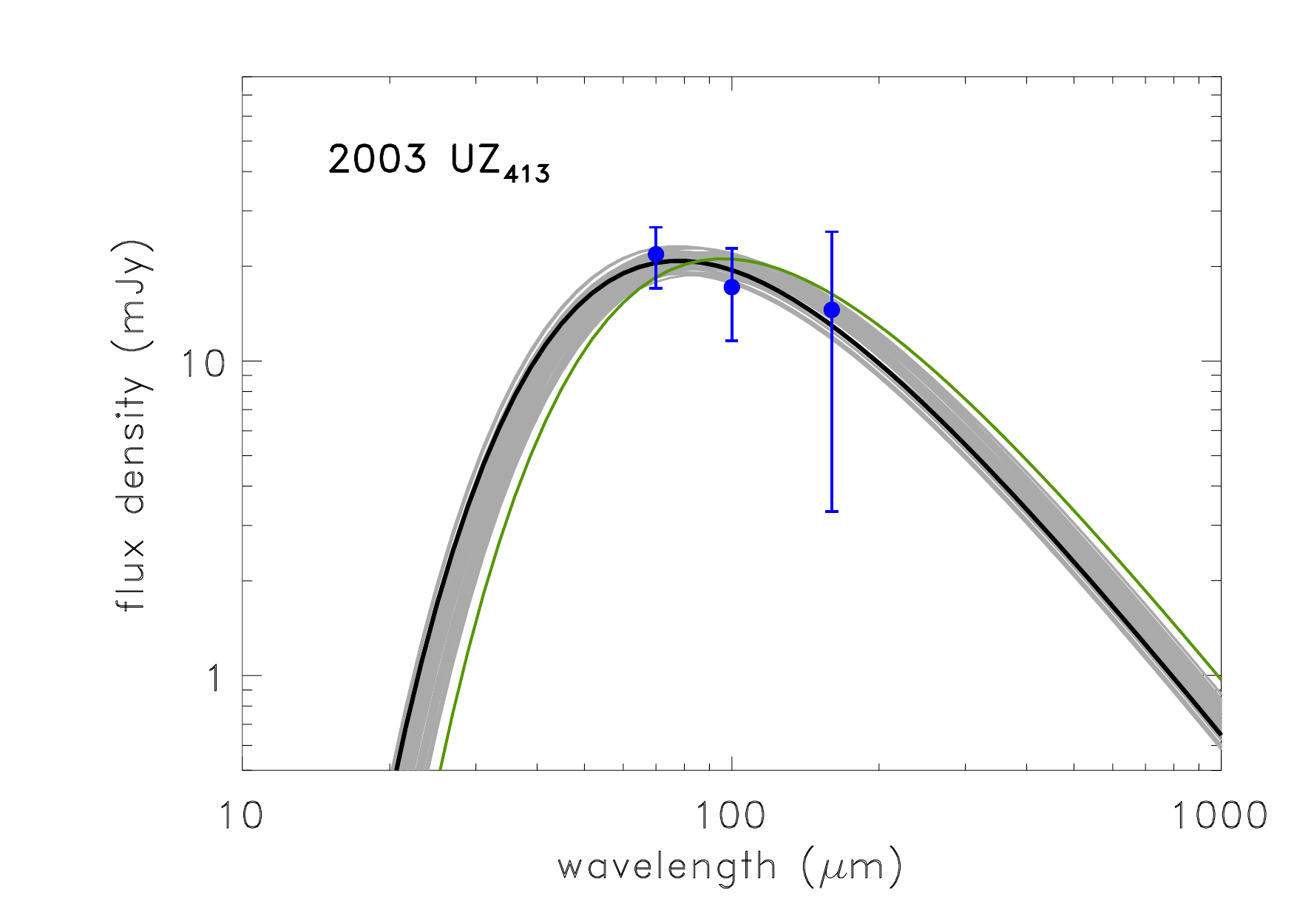}}}
}
\hbox{
\resizebox{5.5cm}{!}{\rotatebox{0}{\includegraphics{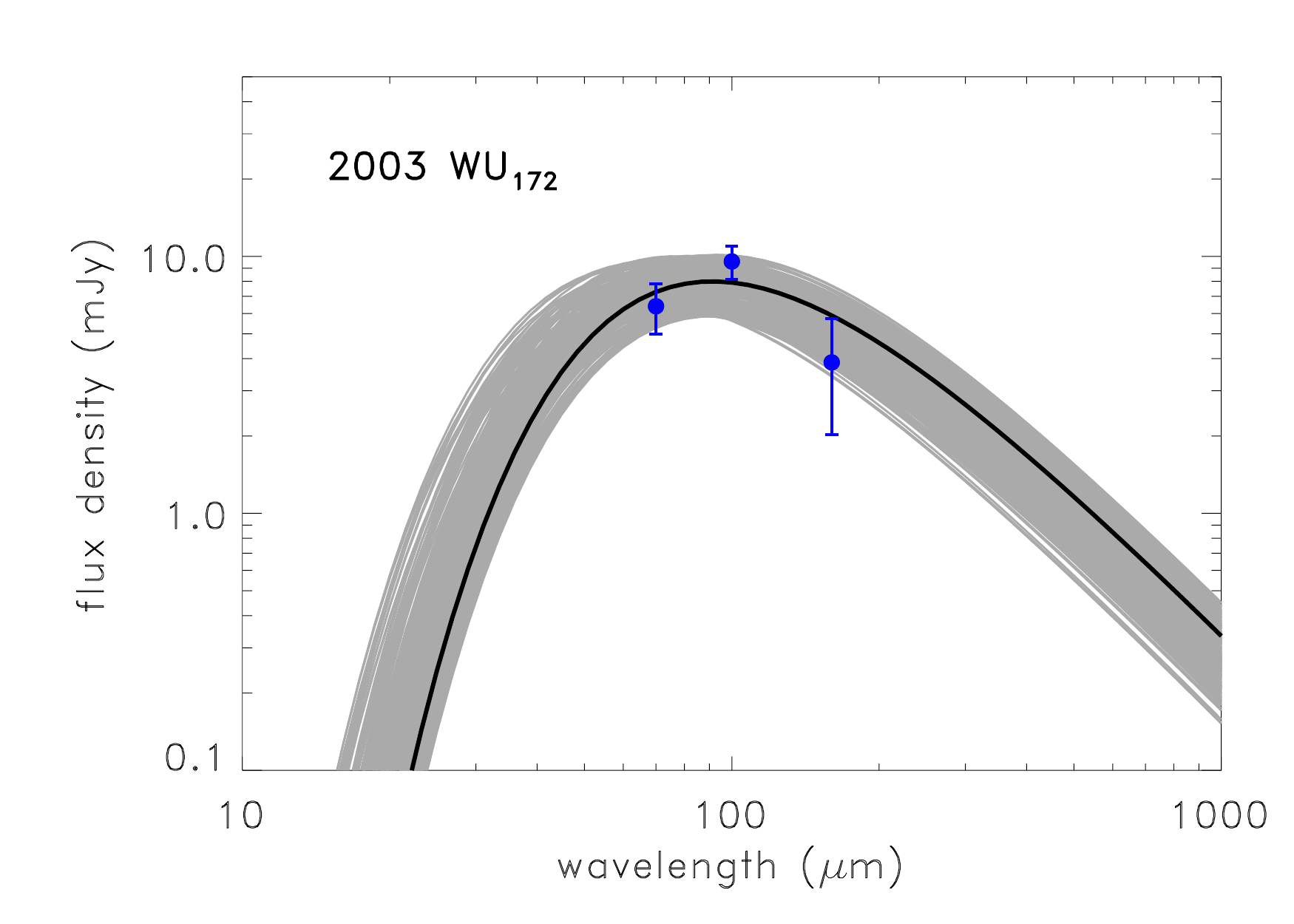}}}
\resizebox{5.5cm}{!}{\rotatebox{0}{\includegraphics{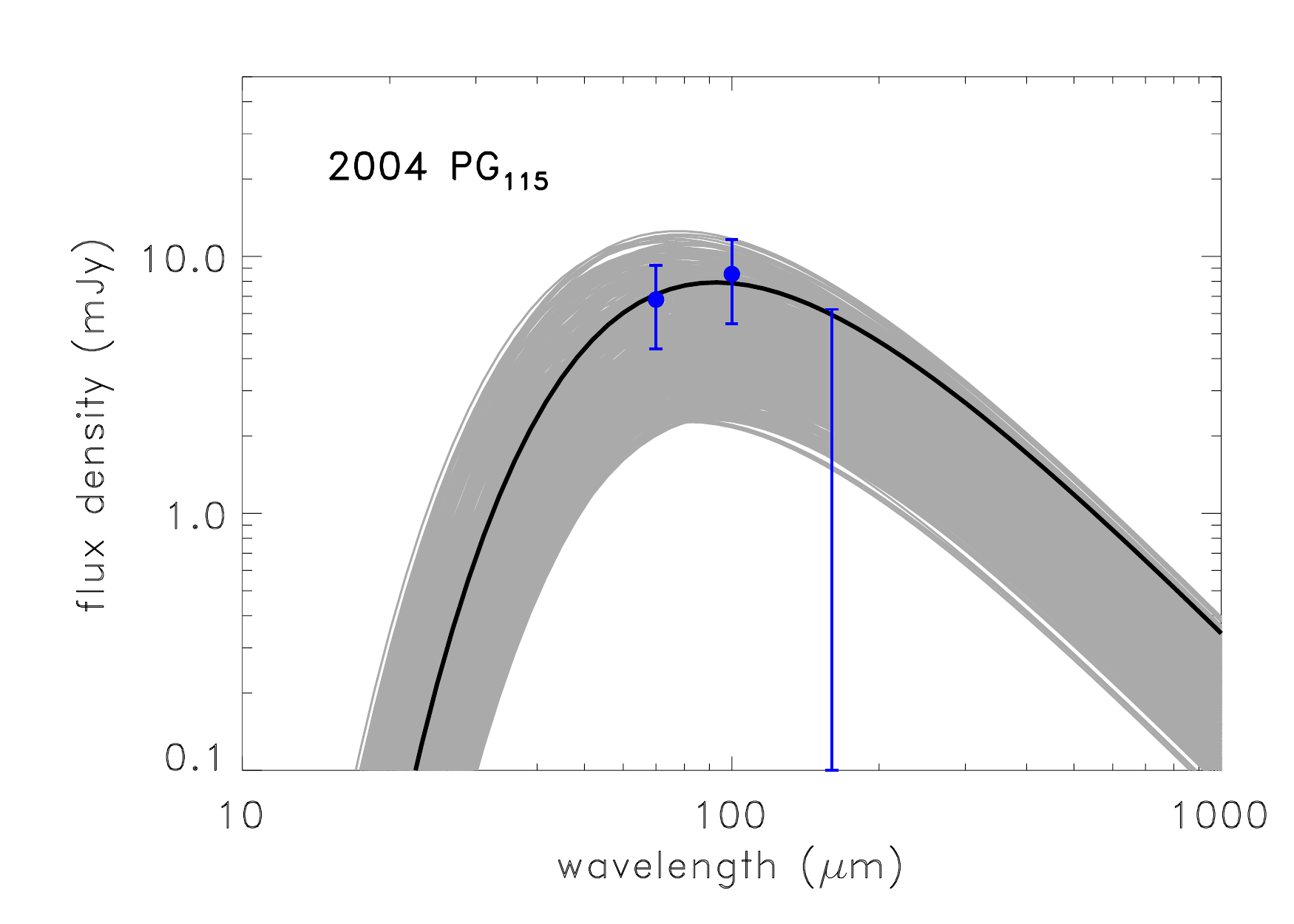}}}
\resizebox{5.5cm}{!}{\rotatebox{0}{\includegraphics{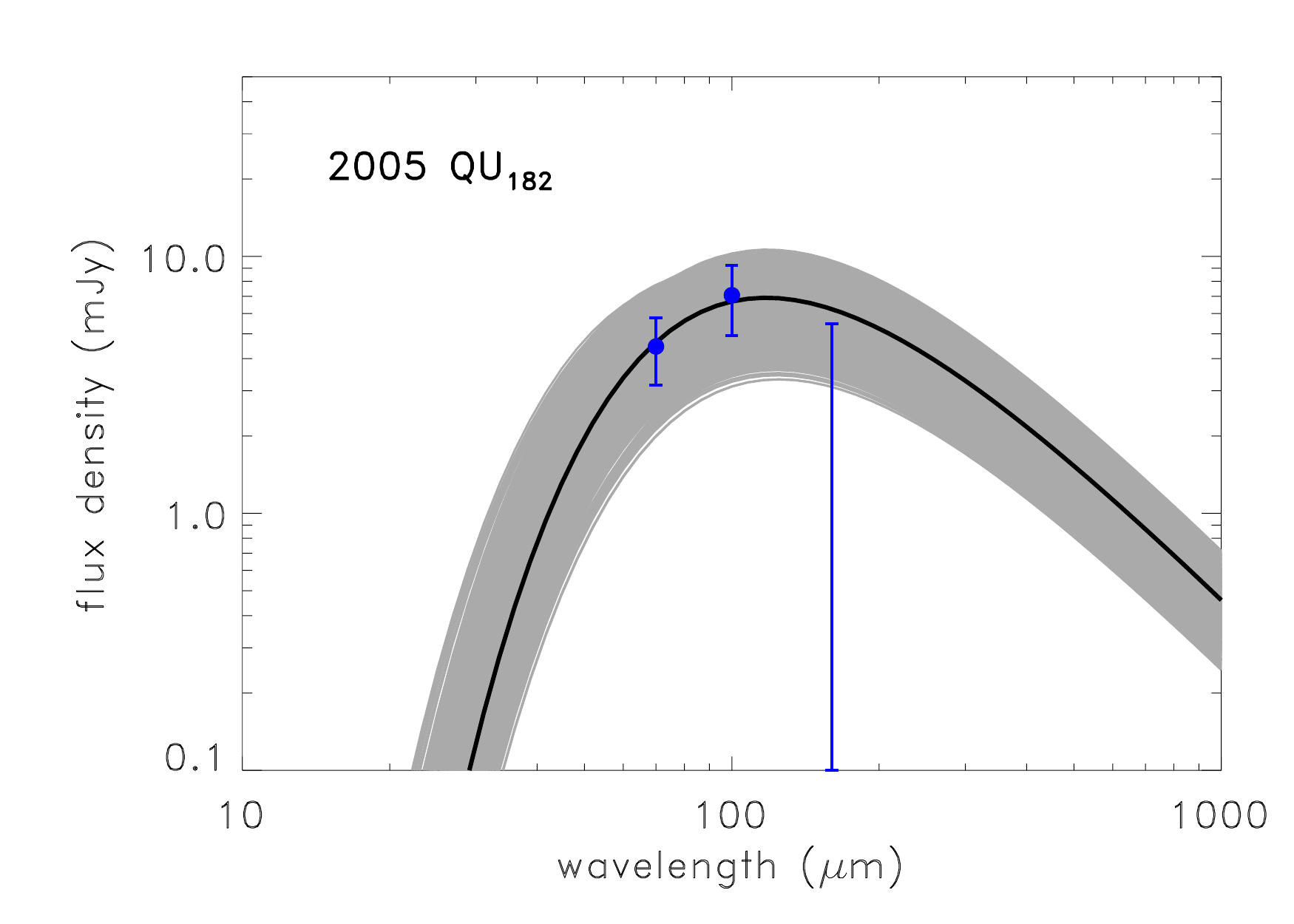}}}
}
\hbox{
\resizebox{5.5cm}{!}{\rotatebox{0}{\includegraphics{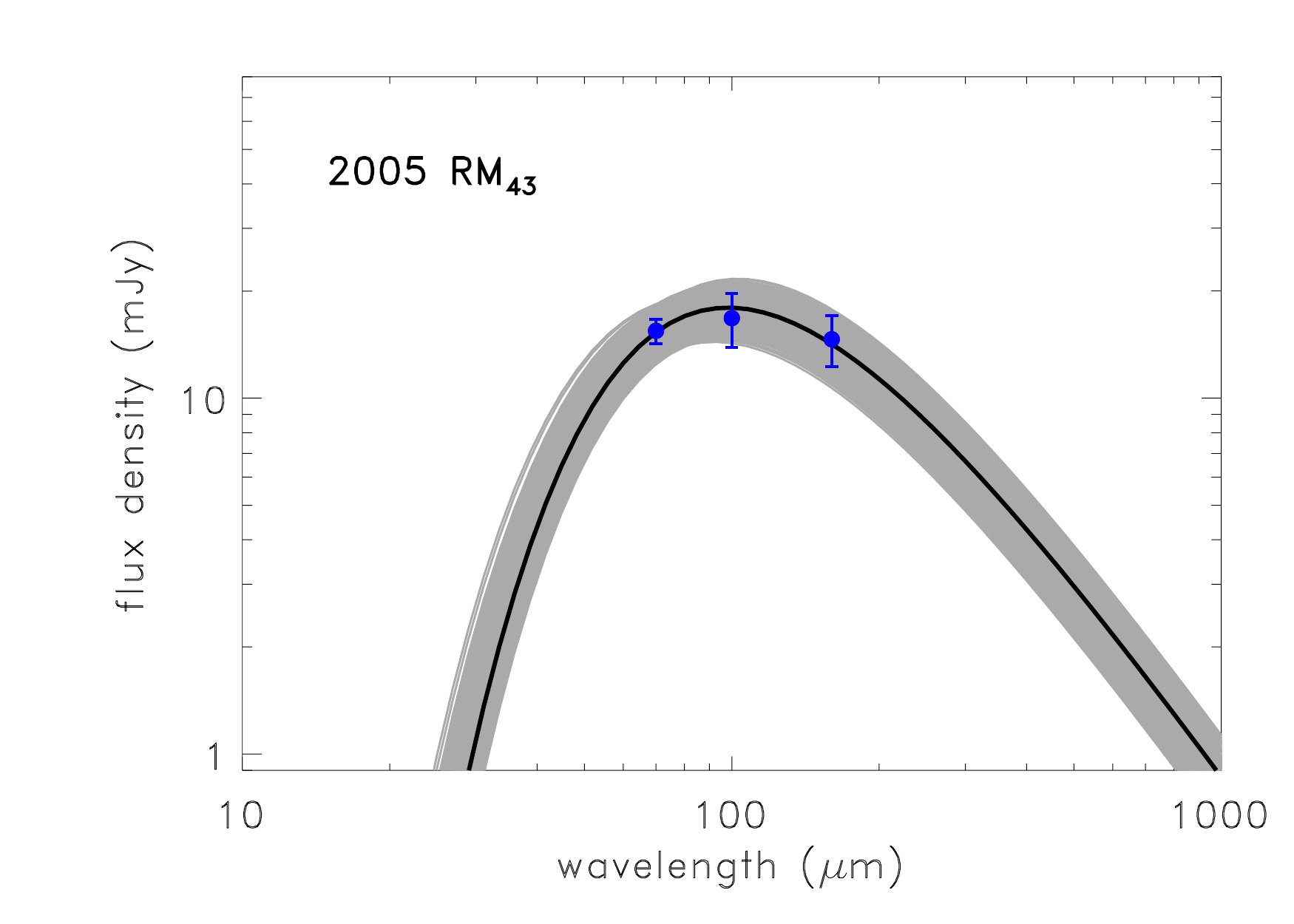}}}
\resizebox{5.5cm}{!}{\rotatebox{0}{\includegraphics{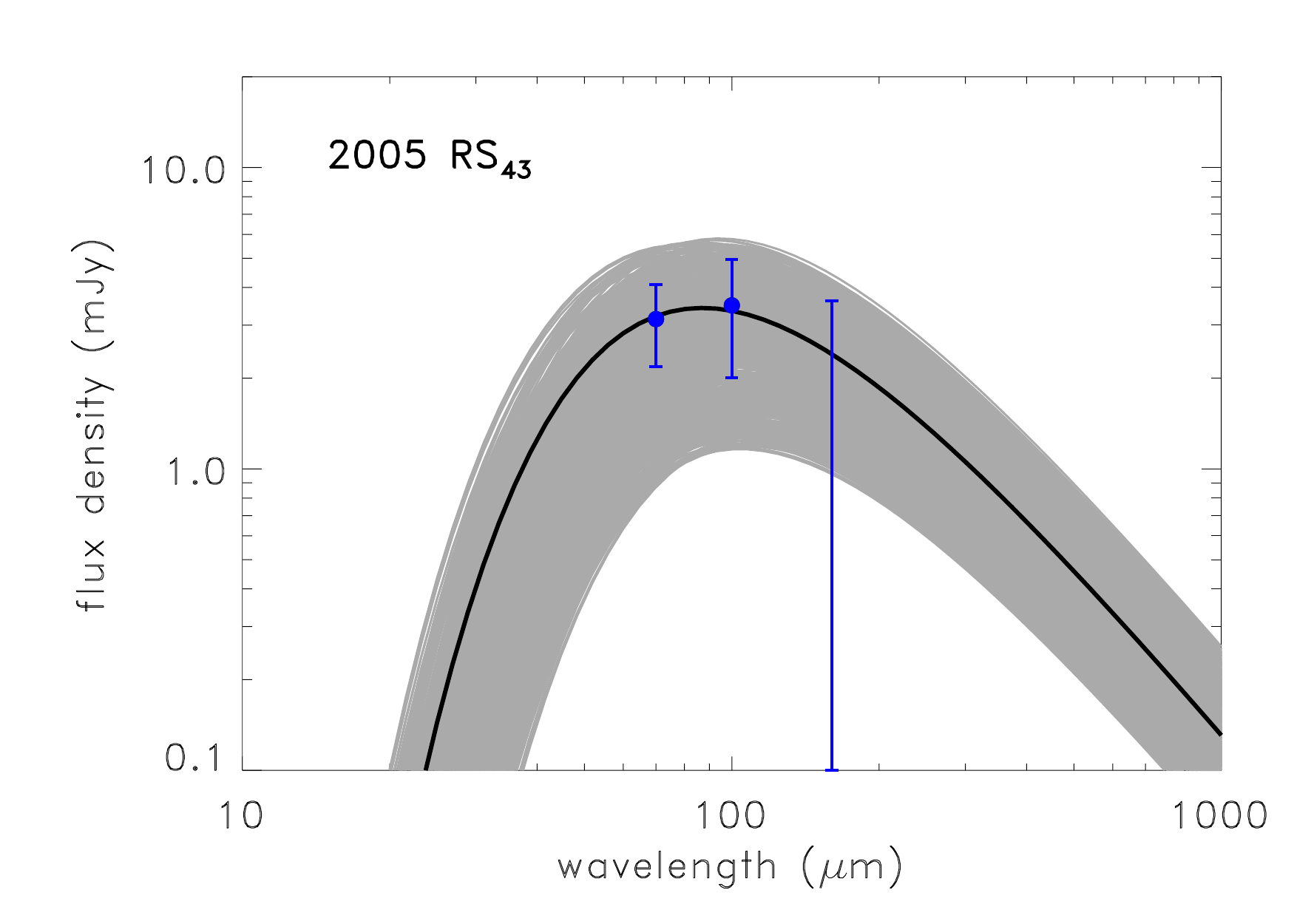}}}
}
\caption{Continuation of Fig.~\ref{fig:seds} \label{fig:seds2}}
\end{figure*}




\subsection{Individual targets \label{sect:individual}}

The NEATM-fits of the individual targets are presented in Fig.~\ref{fig:seds}.

\paragraph{(82075)~2000\,YW$_{134}$} was observed in the Science Verification Phase of the Herschel Space Observatory with the PACS photometer, in chop-nod mode \citep{Muller2010}. Those measurements provided only upper limits of F$_{70}$\,$\leq$\,5\,mJy and F$_{160}$\,$\leq$\,8\,mJy in the 70 and 160\,$\mu$m bands, setting upper limits on the effective diameter, \deff\,$\leq$\,500\,km, and a lower limit of \geomalb\,$\geq$\,0.08 on the geometric albedo. These limits are in agreement with our newly derived, system-integrated values of \deff\,=\,437$^{+118}_{-137}$\,km and \geomalb\,=\,0.13$^{+0.17}_{-0.05}$.
2000\,YW$_{134}$ is a binary, with a brightness difference of $\Delta m$\,=\,1\fm3 between the two components \citep{SN2006}. This provides effective diameters of \deff\,=\,382\,km and 210\,km, assuming equal albedos for the primary and secondary. 

\paragraph{(139775) \qg} is a plutino showing a large amplitude light curve, suggesting that it is a contact binary system \citep{Lacerda2007}. 
Due to the non-detections (upper limits) in all PACS bands, we were only able to derive an upper limit for the equivalent diameter \deff\,$<$\,215\,km for the system and a lower limit for the geometric albedo, \geomalb\,$>$\,0.07. \citet{Lacerda2007} and \citet{Lacerda2011} obtained a solution for the system from multi-epoch visible light curve observations that the density of the system (for both components) is $\rho$\,=\,0.59\,g\,cm$^{-3}$ with a secondary-to-primary mass ratio of q\,=\,0.84, and two triaxial ellipsoids with primary-to-secondary axis ratios of B/A\,=\,0.72, C/A\,=\,0.64,  b/a\,=0.45 and c/a\,=\,0.41, using a Roche model (lowest $\chi^2$ model with lunar-type scattering). Assuming that the two bodies have equal albedos and that the Herschel measurements represent a mean rotational and orbital phase configuration in terms of apparent cross sections of the components, the semi-axes of the ellipsoids are A\,=\,115\,km, B\,=\,82\,km, C\,=\,72\,km, a\,=\,146\,km, b\,=\,66\,km, and c\,=\,60\,km, using the upper limit of \deff\,$<$\,215\,km, from our radiometric solution.

\paragraph{2001\,QR$_{322}$} is the first Neptune Trojan discovered, and it may have a dynamically unstable orbit \citep{Horner2010}. Our measurements suggest a very dark surface (\geomalb\,=\,0.03$\pm 0.02$) and an effective diameter of \deff\,=\,178$^{+61}_{-37}$\,km. 

\paragraph{(42301)~2001\,UR$_{163}$} is one of the reddest objects known in the Solar System \citep[B--R\,=\,2.05$\pm$0.10][]{SS09}. Our floating beaming parameter fit converges to a very low value of $\eta$\,=\,0.5, the lowest limit in our calculations. These extremely low beaming parameters would be expected for a rough surface and low thermal parameters \citep{Spencer1989,Spencer1990}.  
The corresponding albedo and diameter would be  \deff\,=\,261$_{-91}^{+209}$\,km and  \geomalb\,=\,0.43$_{-0.30}^{+0.62}$. We have repeated this calculation by applying a fixed beaming parameter of $\eta$\,=\,1.25, as in previous ``TNOs are Cool'' sample papers (e.g., \citealt{SS12}). This provides a size of  \deff\,=\,367$_{-197}^{+103}$\,km and a geometric albedo of \geomalb\,=\,0.22$_{-0.09}^{+0.83}$.  

\paragraph{(126154)~2001\,YH$_{140}$} was also observed in the Science Verification Phase (SVP) of the Herschel Space Observatory with the PACS photometer, in chop-nod mode \citep{Muller2010} in two bands, 70 and 160\,$\mu$m. The  monochromatic flux densities obtained are F$_{70}'$\,=\,9.8$\pm$2.9\,mJy and F$_{160}'$\,$\leq$\,13\,mJy (i.e., an upper limit in the latter case). The in-band flux densities derived in our present work, F$_{70}$\,=\,5.08$\pm$\,1.40\,mJy and F$_{100}$=\,3.67$\pm$\,1.70\,mJy, are notably lower then suggested by the earlier measurements, and we provide a more strict upper limit of F$_{160}$\,$\leq$\,1.95\,mJy (1$\sigma$). The high flux densities obtained in the SVP measurements may be a consequence of the weak reliability of the chop-nod observing mode that was superseded by the scan-map mode in the later phases of the Herschel mission \citep{Nielbock}, and are not necessarily true flux density changes between the two observational epochs. Based on our new radiometric fits 2001\,YH$_{140}$ is smaller and brighter, \deff\,=\,252$_{-52}^{+148}$\,km and \geomalb\,=\,0.14$_{-0.09}^{+0.14}$, than that derived from the Herschel SVP measurements (D\,$\approx$\,350\,km and \geomalb\,$\approx$\,0.08).

\paragraph{2003\,UZ$_{413}$} is a plutino with a rotation period of P\,=\,4.13$\pm$0.05\,h and a light curve amplitude of $\Delta m$\,=\,0.13$\pm$0.03\,mag \citep{Perna2009}. The observed fast rotation and the light curve amplitude might be compatible with a Jacobi ellipsoid with $a$/$b$\,$\approx$\,1.13; however, this would require a high density of $\rho$\,=\,2.3--3.0\,\gcc, a peculiarly high value among trans-Neptunian objects. 
For 2003\,UZ$_{413}$, the floating $\eta$ NEATM fit in our radiometric analysis cannot provide a well-defined solution ($\eta$ converges to the minimum allowed value of 0.5). By applying a fixed beaming parameter of $\eta$\,=\,1.25, we derived an effective diameter of \deff\,=\,650$^{+1}_{-175}$\,km and a geometric albedo of \geomalb\,=\,0.08$^{+0.08}_{0.01}$. 

\paragraph{(145451)~2005\,RM$_{43}$} is a detached object that was also observed by \citet{Perna2009}, who obtained a rotation period of P\,=\,9.00$\pm$0.06\,h and a light curve amplitude of $\Delta m$\,=\,0.12$\pm$0.05\,mag. A Jacobi ellipsoid model provides a density estimate of $\rho$\,$\approx$\,0.55\gcc. According to our radiometric fit, the diameter is \deff\,=\,524$_{-103}^{+96}$\,km with a geometric albedo of \geomalb\,=\,0.10$^{+0.06}_{-0.03}$. 

\paragraph{(40314)~1999\,KR$_{16}$ and 2005\,QU$_{182}$} were analyzed previously in \citet{SS12}, but the Herschel/PACS 100\,$\mu$m data were inconsistent (too low) with the flux densities obtained in the 70 and 160\,$\mu$m bands, and we therefore  reanalyzed these measurements in the present work. 

As discussed in Sect.~2.1, in our reduction we used double-differential images instead of supersky-subtracted images, as in \citet{SS12}, to minimize the effect of nearby background sources. 
In the case of 1999\,KR$_{16}$ \citet{Vilenius2018} reanalyzed the PACS measurements and found that there was a bright background source near the target at one of the epochs; due to this contamination they decided  not to use the visit~2 measurements (i.e., their photometry is based on a single-epoch measurement, strongly affected by the background). 

For 1999\,KR$_{16}$ \citet{SS12} obtained the monochromatic flux densities of  F$_{70}'$\,=\,5.7$\pm$0.7\,mJy, F$_{100}'$\,=\,3.5$\pm$1.0\,mJy, and F$_{160}'$\,=\,4.6$\pm$2.2\,mJy. Our in-band flux densities are F$_{70}$\,=\,7.61$\pm$1.44\,mJy, F$_{100}$\,=\,5.21$\pm$2.29\,mJy, and F$_{160}$\,$<$\,1.98\,mJy. The fixed beaming parameter fit of $\eta$\,=\,1.2 in \citet{SS12} provided \deff\,=\,254$\pm$37\,km and \geomalb\,=\,0.20$^{+0.07}_{-0.05}$. This is similar to our floating $\eta$ fit, \deff\,=\,202$^{+108}_{-36}$\,km and \geomalb\,=\,0.31$^{+0.16}_{-0.18}$. However, in this latter case the best-fit beaming parameter is $\eta$\,=\,0.53$^{+1.84}_{-0.02}$, an extremely low value. 
For 2005\,QU$_{182}$ \citet{SS12} obtained F$_{70}'$\,=\,4.5$\pm$0.9\,mJy, F$_{100}'$\,=\,2.5$\pm$1.1\,mJy, and F$_{160}'$\,=\,8.4$\pm$3.0\,mJy, providing \deff\,=\,416$\pm$73\,km and \geomalb\,=\,0.33$^{+0.16}_{-0.11}$ using a fixed $\eta$\,=\,1.2 fit. In our photometry using the double-differential products, the monochromatic flux densities are F$_{70}$\,=\,4.41$\pm$1.28\,mJy, F$_{100}$\,=\,6.94$\pm$2.11\,mJy, and F$_{160}$\,$<$\,2.73\,mJy, and the corresponding effective diameter and geometric albedo are \deff\,=\,584$^{+155}_{-144}$\,km and \geomalb\,=\,0.13$^{+0.12}_{-0.05}$, with a best-fit beaming parameter of $\eta$\,=\,2.08$^{+0.42}_{-1.37}$ (i.e., a notably larger size and lower albedo). 

\begin{figure}[!ht]
\includegraphics[width=0.405\textwidth]{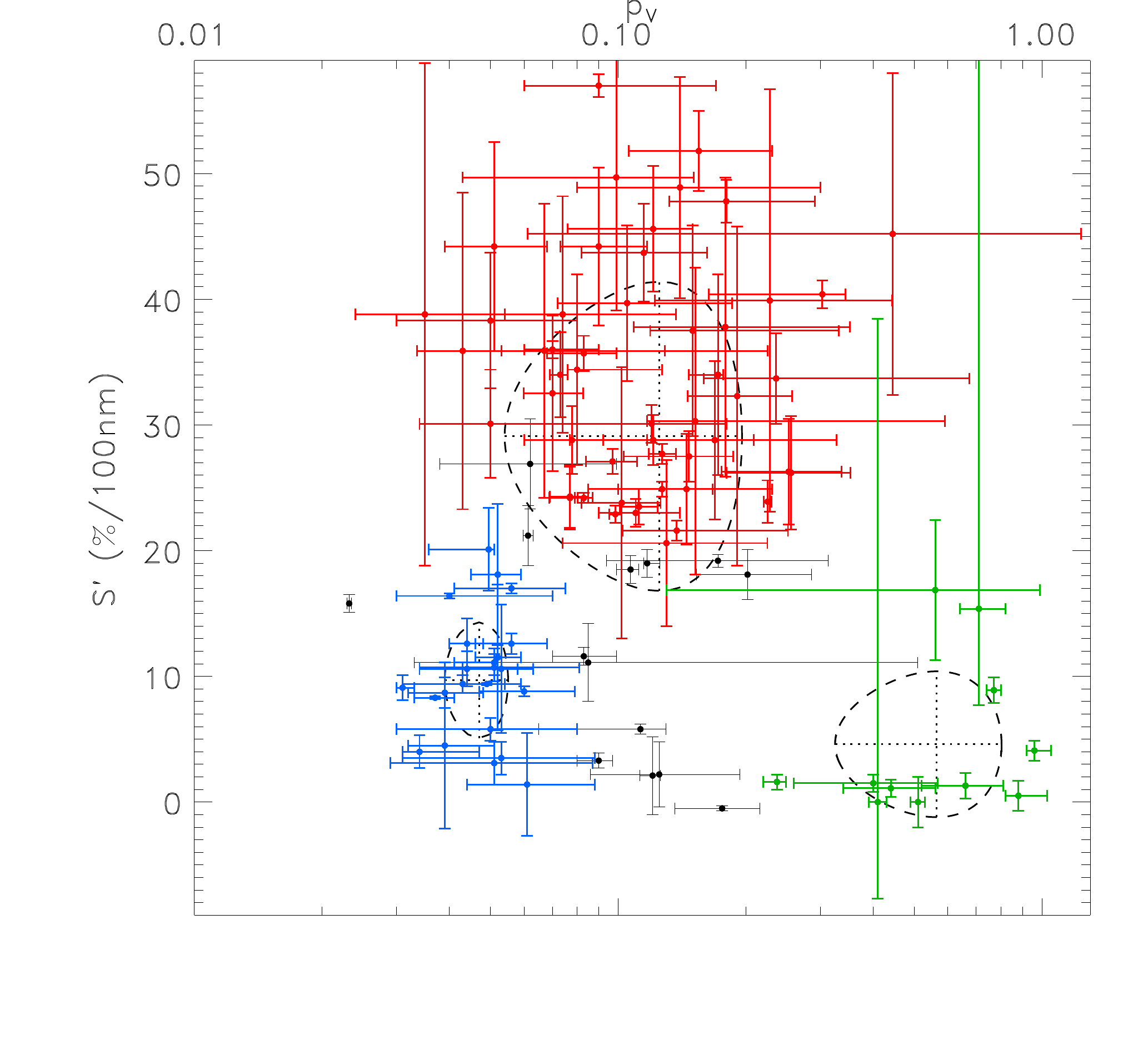}

\includegraphics[width=0.405\textwidth]{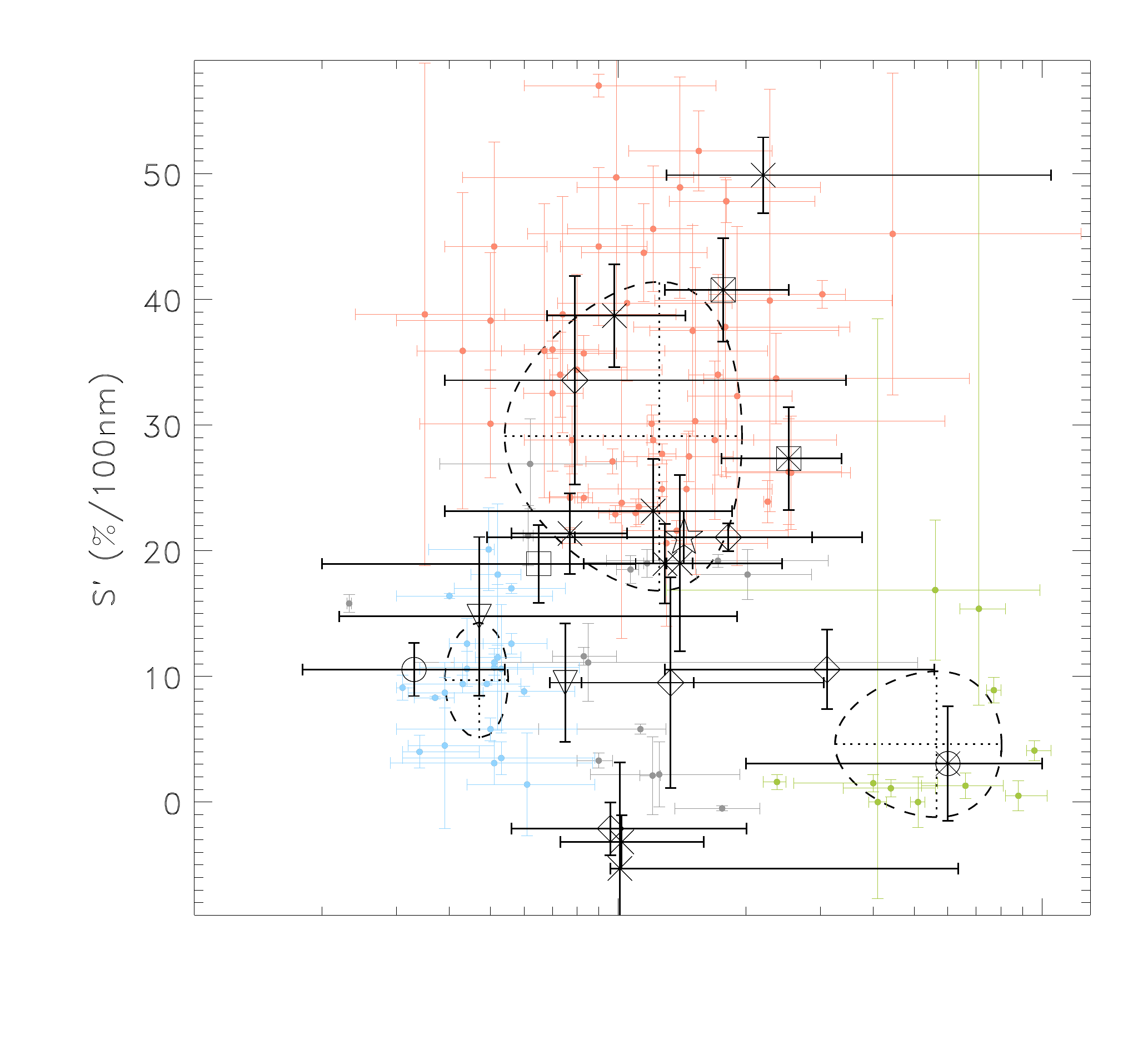}

\includegraphics[width=0.405\textwidth]{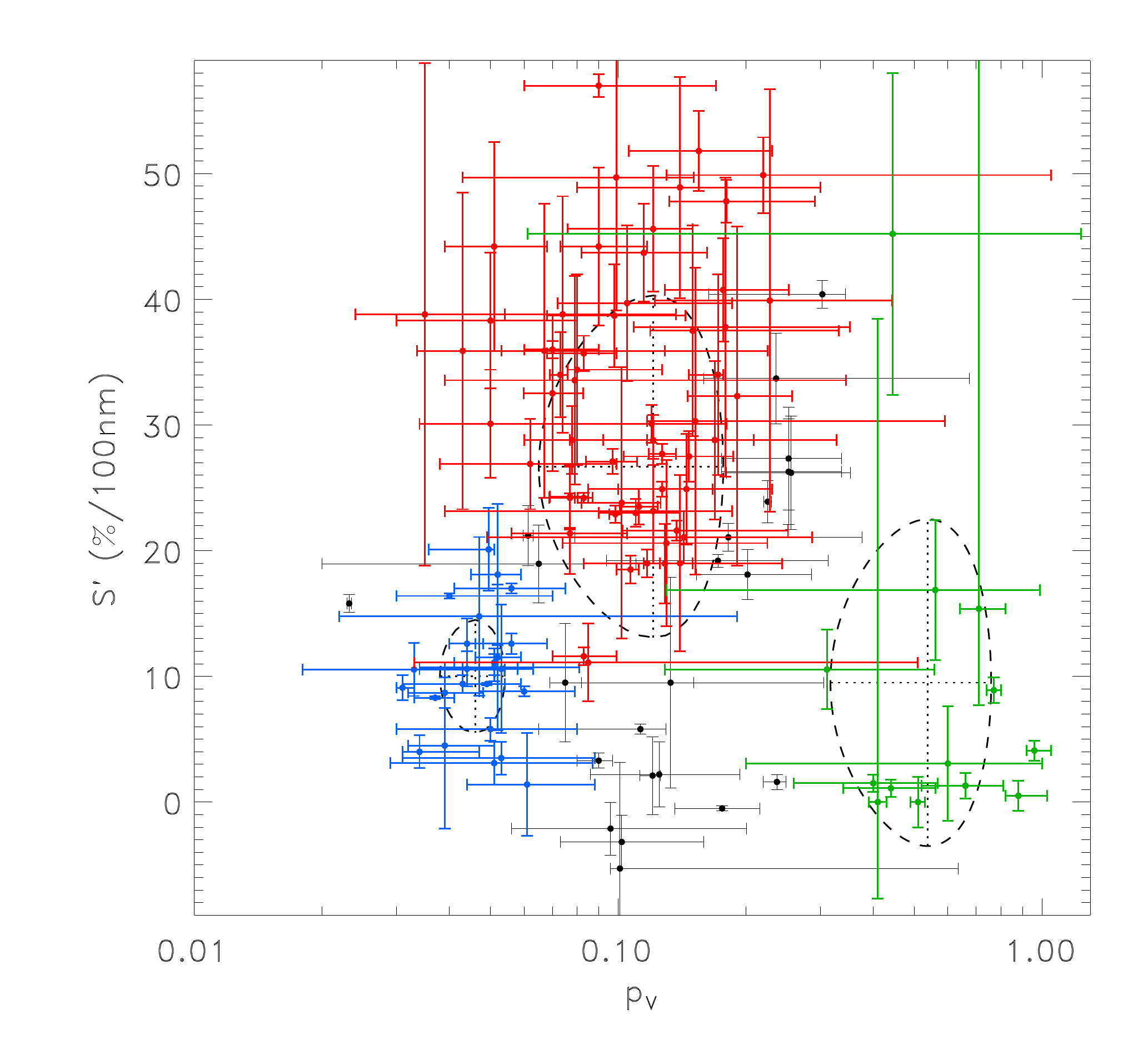}
\caption{Albedo vs. visible color, quantified by the spectral slope $S'$. 
{\it Upper panel}: TNOs from \cite{Lacerda2014} using updated albedo and spectral slope values in some specific cases. The three clusters identified are indicated in  blue, red, and green. Ellipses (distorted due to the logarithmic \geomalb{} scale) with crosshairs indicating the centers and variance values of the clusters in the albedo--spectral slope plane. Objects indicated by black symbols have ambiguous identifications.  
{\it Middle panel}: Same as the top panel (TNOs from Lacerda et al., 2014), but with the new objects of the present paper overplotted. The following symbols were used: 
$\triangledown$ -- plutinos;
$\boxtimes$ -- detached objects;
$\times$ -- scattered disk objects; 
$\Diamond$ -- outer resonants;
$\Circle$ -- inner resonants;
$\star$ -- middle resonants;
$\Box$ -- 1999\,CD$_{158}$ \citep{Vilenius2018}; 
$\otimes$ -- 1996\,TO$_{66}$ \citep{Vilenius2018}.
{\it Bottom panel}: Clusters in the albedo vs. spectral slope plot using all 120 objects, including those from Lacerda et al. (2014) and the new targets from the present paper.   }
\label{fig:slope_pv}
\end{figure}
\section{Discussion}

\subsection{Thermal intertia estimates \label{sect:thermalinertia}}

\citet{Lellouch2013} estimated thermal inertia from the $\eta$ values derived and obtained a decrease in $\Gamma$ from $\sim$5\,\tiunit to $\sim$2\,\tiunit\, from $\sim$25\,AU to heliocentric distances beyond 40\,AU. Following a similar approach we also estimated the thermal inertia values for our targets. The phase integral $q$ was calculated applying both the standard \citet{Bowell1989} formula with a G\,=\,0.15 slope parameter, and also using the geometric albedo dependent formula by \citet{Brucker2009}. 
These two methods resulted in subsolar temperatures with negligible differences (typically $\lesssim$1\,K, the largest difference is $\sim$4\,K for the high albedo target 2002\,CY$_{224}$) and with a negligible effect on the  final results. We assumed a uniform P\,=\,8\,h rotation period for all targets. We used two subsolar latitudes, $\beta_{ss}$\,=\,32.7\degr{} (average values of random orientation spin axis directions) and also $\beta_{ss}$\,=\,0\degr{}, as some specific (high, $\eta$\,$\geq$\,2.0) beaming parameter values could not be reproduced with $\beta_{ss}$\,=\,32.7\degr. We allowed a roughness from $s$\,=\,0 to 60\degr{} (zero to high roughness). 
The thermal inertias obtained were similar in the two $\beta_{ss}$ cases for a specific target, but showed a wide range of possible values, partly due to the typically large uncertainties in the $\eta$ determination, and the ambiguity that different combinations of $\Theta$, $\beta_{ss}$, and $s$ can result in the same beaming parameter at the end. The median values for our sample were 1.9\,\tiunit\, for $\beta_{ss}$\,=\,0\degr{} and {\bf 3.1\,\tiunit\,} for $\beta_{ss}$\,=\,32.7\degr{}; for the same target (albedo and heliocentric distance) a higher $\beta_{ss}$ value results in a higher $\Gamma$, as a higher $\Theta$ is needed to obtain the same beaming parameter (see fig.~4 in Lellouch et al., 2013). These thermal inertias are compatible with those obtained by \citet{Lellouch2013} for the same heliocentric distance range. 

\subsection{Albedo-color diagram}
\label{sec:pv_colour}

As was demonstrated in \citet{Lacerda2014} objects in the trans-Neptunian region can be divided into two main clusters based on their geometric albedo and visible color properties. We repeated this analysis using the {\sl mclust} package in {\sl R} \citep{Fraley}. 

We used two main datasets. The first  is the \citet{Lacerda2014} sample using mainly the geometric albedo and spectral slope data derived from the original papers, but using new albedo and color values in the   cases when they were updated according to the latest reduction of the Herschel and Spitzer data -- e.g., 2007\,OR$_{10}$~\citep{Pal2016} and 2007\,UK$_{126}$~\citep{Schindler2017} -- or by new occultation measurements \citep[Haumea,][]{Ortiz2017,Muller2018b}. 
New targets given in the present paper are not included in this dataset. 

The second dataset contains {\it all} objects, i.e., the Lacerda et al. (2014) sample and also our new targets, and two additional targets, 1996\,TO$_{66}$ and 1999\,CD$_{158}$ from \citep{Vilenius2018}. 1996\,TO$_{66}$ is considered to be a Haumea collisional family member, while 1999\,CD$_{158}$ is thought to be dynamical interloper with this family. We also considered the relatively wide ranges obtained for the albedos and colors from the disk-resolved measurements of Pluto and Charon \citep{Grundy2016} when drawing their data points in the corresponding figures. 

In both cases we run the {\sl MClust} package for a multiple number of cluster components assumed and also using different data models. 
The results are presented in Fig.~\ref{fig:slope_pv} and summarized in Table~\ref{table:clusters}. In the case of the first dataset (Lacerda et al., 2014 sample) the smallest residuals were obtained using an ellipsoidal, varying volume and shape data model (VVI) with three groups. These clusters are very closely identical with the clusters obtained by Lacerda et al. (2014), i.e., we can identify a dark-neutral group (DN, blue symbols in Fig.~\ref{fig:slope_pv}) and a bright-red group (BR, red symbols), and a third group at high albedos and nearly solar colors. This last group is not defined explicitly by Lacerda et al. (2014) as the related objects are essentially excluded from that analysis;  these objects are the brightest dwarf planets (Pluto, Eris, Makemake) and the members of the Haumea collisional family. As we left these   objects in the dataset, a third, ``bright-neutral'' (BN) group could be identified (Fig.~\ref{fig:slope_pv}, left panel, green symbols). We note that the objects with uncertain cluster identification are indicated by black symbols in the figure (located typically between two clusters). 

We plotted the new targets over the clusters of the original sample in the middle panel of Fig.~\ref{fig:slope_pv}. These targets extend the spectral slope range to smaller values (close to solar) even in the p$_V$\,$>$\,0.1 region, where there were few objects in the original sample. Including the new targets in the cluster modeling results in somewhat different clusters (Fig.~\ref{fig:slope_pv}, right panel), but the main structure (the number, location, and extension of the groups) remains basically unchanged (see Table~\ref{table:clusters}). The groups we identified both in the original and in the full sample match the clusters identified by \citet{Lacerda2014} very closely. 

While most of our new objects follow the previously identified main groups, there is an important new feature. In the \citet{Lacerda2014} analysis the albedo--color range of 0.10\,$\lesssim$\,\geomalb\,$\lesssim$0.40 and 5\,$\lesssim$\,$S'$\,$\lesssim$15\, [\%/100\,nm] remained empty, in our sample we have three objects, 2000\,YW$_{134}$, 1995\,QY$_{9}$, and 2005\,RS$_{43}$ that fall here. While many inner Solar System objects have surfaces matching the albedos and colors of our dark and neutral group \citep{Lacerda2014}, objects with these moderate albedos and slightly red colors cannot be found even in the inner Solar System beyond the 5:3 resonance with Jupiter (including Hildas and Jupiter Trojans), only at heliocentric distances of $\lesssim$3.5\,au \citep{DeMeo2014}. These objects seem to lack the red material that is common among most trans-Neptunian objects in this albedo range. We note that the Neptune irregular satellite Nereid has albedo and color similar to those of these objects \citep{Kiss2016}. 
Targets with nearly solar colors and medium albedos (0.08\,$\lesssim$\,\geomalb\,$\lesssim$\,0.30) all have ambiguous identifications in this clustering scheme.

\begin{table*}[ht!]
    \centering
    \begin{tabular}{ccc|ccc|ccc}
    \hline
     \multicolumn{3}{c|}{Lacerda et al. (2014)} & \multicolumn{3}{c|}{Original sample} &                     \multicolumn{3}{c}{All targets} \\
    \multicolumn{3}{c|}{\sl Mathematica/FindClusters} & \multicolumn{3}{c|}{\sl R/MClust, VVI} &                     \multicolumn{3}{c}{\sl R/MClust, VVI} \\
    \hline
     p$_V$ & $S'$ & ID & p$_V$ & $S'$ & ID & p$_V$ & $S'$ & ID \\
    \hline
$\sim$0.05 & $\sim$10 & DN & 0.047$\pm$0.008 &  9.7$\pm$4.6  & DN & 0.046$\pm$0.008 & 10.0$\pm$4.5 & DN \\
$\sim$0.15 & $\sim$35 & BR & 0.125$\pm$0.071 & 29.1$\pm$12.3 & BR & 0.121$\pm$0.056 & 26.7$\pm$13.6 & BR \\
$>$0.30 & $\sim$0     & BN & 0.564$\pm$0.239 &  4.6$\pm$5.8  & BN & 0.538$\pm$0.221 & 9.5$\pm$13.0 & BN \\
     \hline
    \end{tabular}
    \caption{Clusters identified by Lacerda et al. (2014, Col. 1), using the Lacerda et al. (2014) sample with updates (this work, Col. 2), and using all targets (this work, Col. 3). Clusters are characterized here by their mean geometric albedo (p$_V$) and spectral slope ($S'$, \%/1000\,$\AA$), and the standard deviations using an ellipsoidal varying volume and shape data model. In the second row we also list the tools used to identify the clusters.}
    \label{table:clusters}
\end{table*}

\begin{figure*}
\hbox{\includegraphics[width=0.5\textwidth]{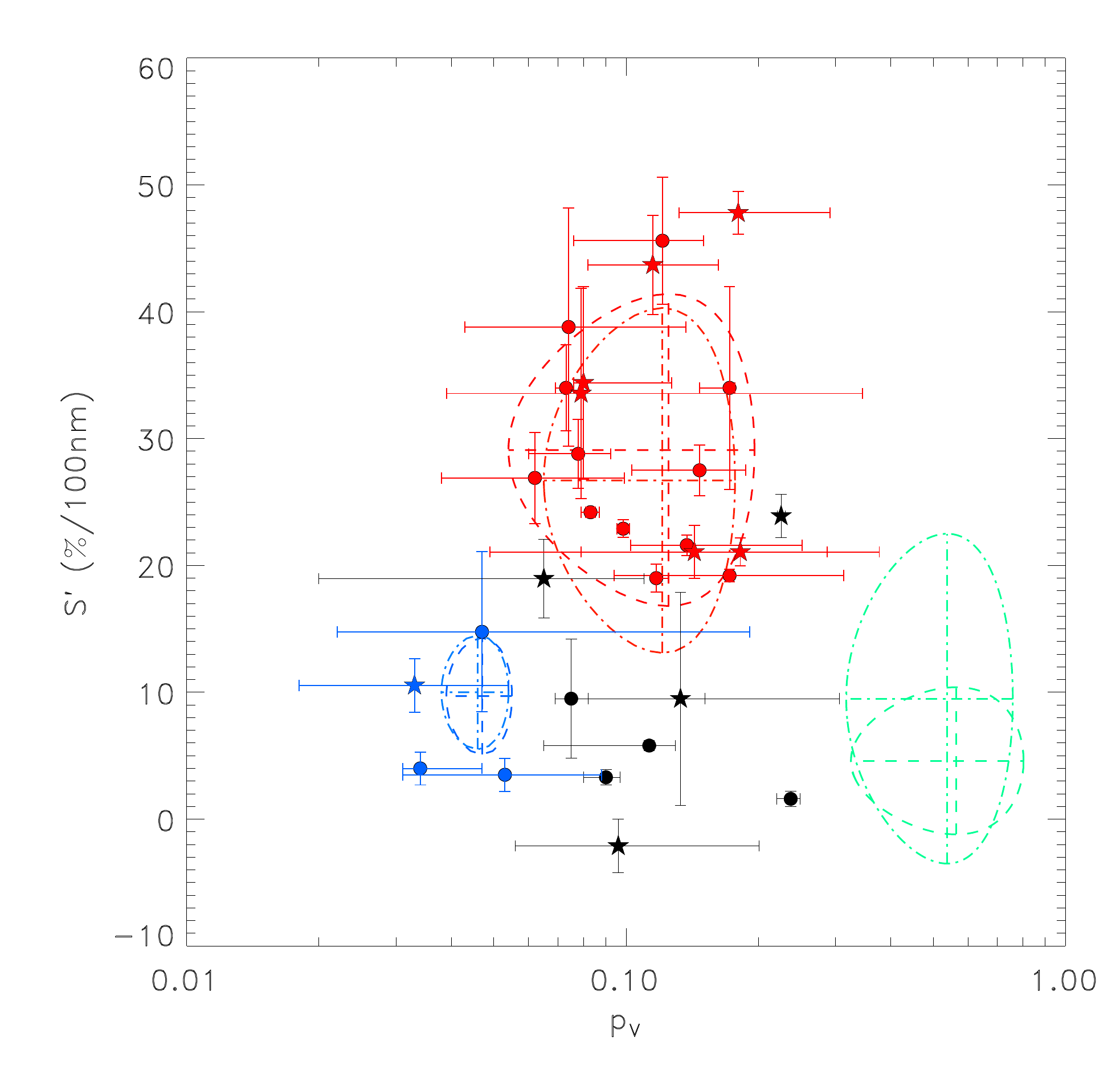}
\includegraphics[width=0.5\textwidth]{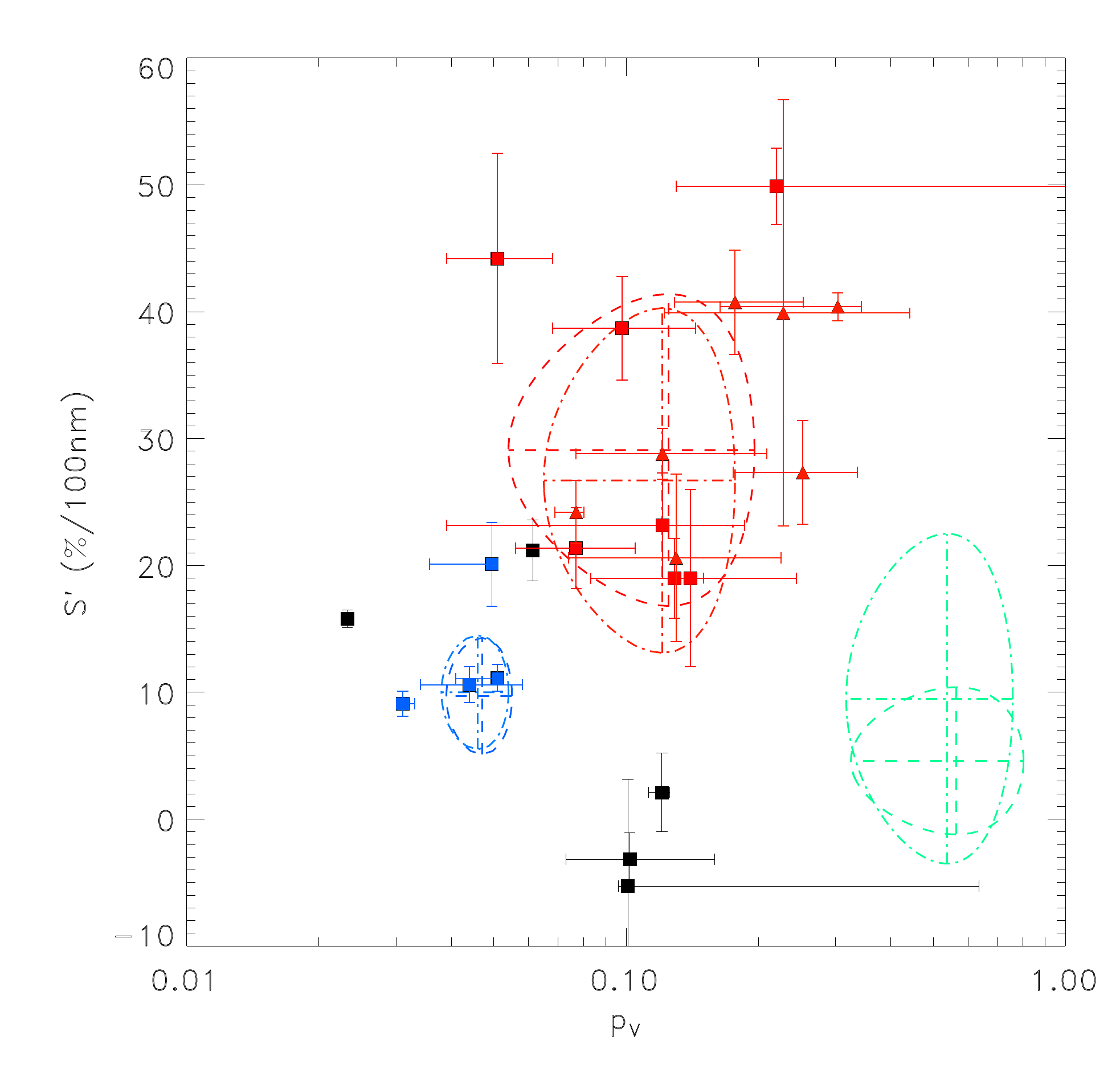}}
\caption{Spectral slope vs. geometric albedo for resonant (left), and scattered disk and detached objects (right). Regions delineated with  dashed and dash-dotted lines correspond to the 1\,$\sigma$ contours of the clusters identified in our analysis using the original and full samples, respectively. 
Filled circles and stars mark plutinos (3:2 resonance) and other resonant objects, respectively, in the left panel. Filled squares and triangles  indicate scattered disk and detached objects in the right panel. Black symbols indicate objects with ambiguous cluster assignments. }
\label{fig:clusters4}
\end{figure*}

We also found that  the objects 2004\,PG$_{115}$ and 2005\,RM$_{43}$, although they are likely associated with the DN group, are located so far from the group center with their negative spectral slopes and \geomalb\,$\approx$\,0.10 albedos that their group assignment can be considered  ambiguous. \citet{Lacerda2014} found that objects in relic populations, i.e.,  those that formed and remained in the outer Solar System (cold classical KBOs and objects in the outer resonances), possess exclusively bright red surfaces, while populations with dynamically more complex origin (Centaurs, scattered disk objects, hot classical KBOs, plutinos) show mixed surfaces. Our new targets follow this general picture; for example, new scattered disk objects can be found both in the DN and the BR groups, while detached objects and targets from the outer resonances can exclusively be found among BR objects.

\subsection{Impact on size distributions}

Size distribution is an essential characteristic of small body populations, providing information on the collisional history. As diameters of small bodies cannot be obtained directly in most cases, it is difficult to obtain a reliable size distribution. Usually it is necessary to rely on the absolute magnitude distribution that can be derived directly from the observations of these objects \citep{Gladman2001}. Typically,  a single, mean albedo value known from other measurements can be used (e.g., radiometry or occultations) to convert between the diameter and the absolute magnitude in the corresponding distributions.  \citet{Mommert12}, among others,  used an average albedo of 0.08$\pm$0.03 to convert from absolute magnitude to effective diameter and to derive the size distribution of plutinos. However, as we have shown above, trans-Neptunian populations, including scattered disk and resonant objects show a bimodal albedo and color distribution; therefore, the application of a bimodal albedo (e.g., using the color as a proxy for the mean albedo) might be more appropriate. 

In Fig.~\ref{fig:clusters4} we drew the plutinos, objects in other (not 3:2) resonances, and scattered disk and detached objects over the main groups identified by our cluster analysis. Bimodal and color-dependent albedo is clearly the case for scattered disk and detached objects (right panel). Here the objects in the specific groups, \geomalb(SDO,DN)\,=\,0.044$\pm$0.008 and \geomalb(SDO,BR)\,=\,0.119$\pm$0.054, follow the general albedos of the larger samples (Table~\ref{table:clusters}). We note that there is no detached object in the DN group. The three objects with very low spectral slope values ($S'$\,$\lesssim$\,0, 2000\,YW$_{134}$, 1995\,QY$_{9}$ and 2005\,RS$_{43}$) seem to have albedos similar to the BR group, \geomalb\,$\approx$\,0.12. The case is different for the resonant objects, including the plutinos, due to the presence of several new objects in the range 0.10\,$\lesssim$\,\geomalb\,$\lesssim$0.40 and 5\,$\lesssim$\,$S'$\,$\lesssim$15\, [\%/100\,nm] (see the discussion above), making the color-to-albedo assignment ambiguous. It is, however, still the case for the BR group that their colors correspond to a relatively well-defined albedo, \geomalb(RTNO,BR)\,=\,0.12$\pm$0.04.


\begin{figure}
\includegraphics[width=0.5\textwidth]{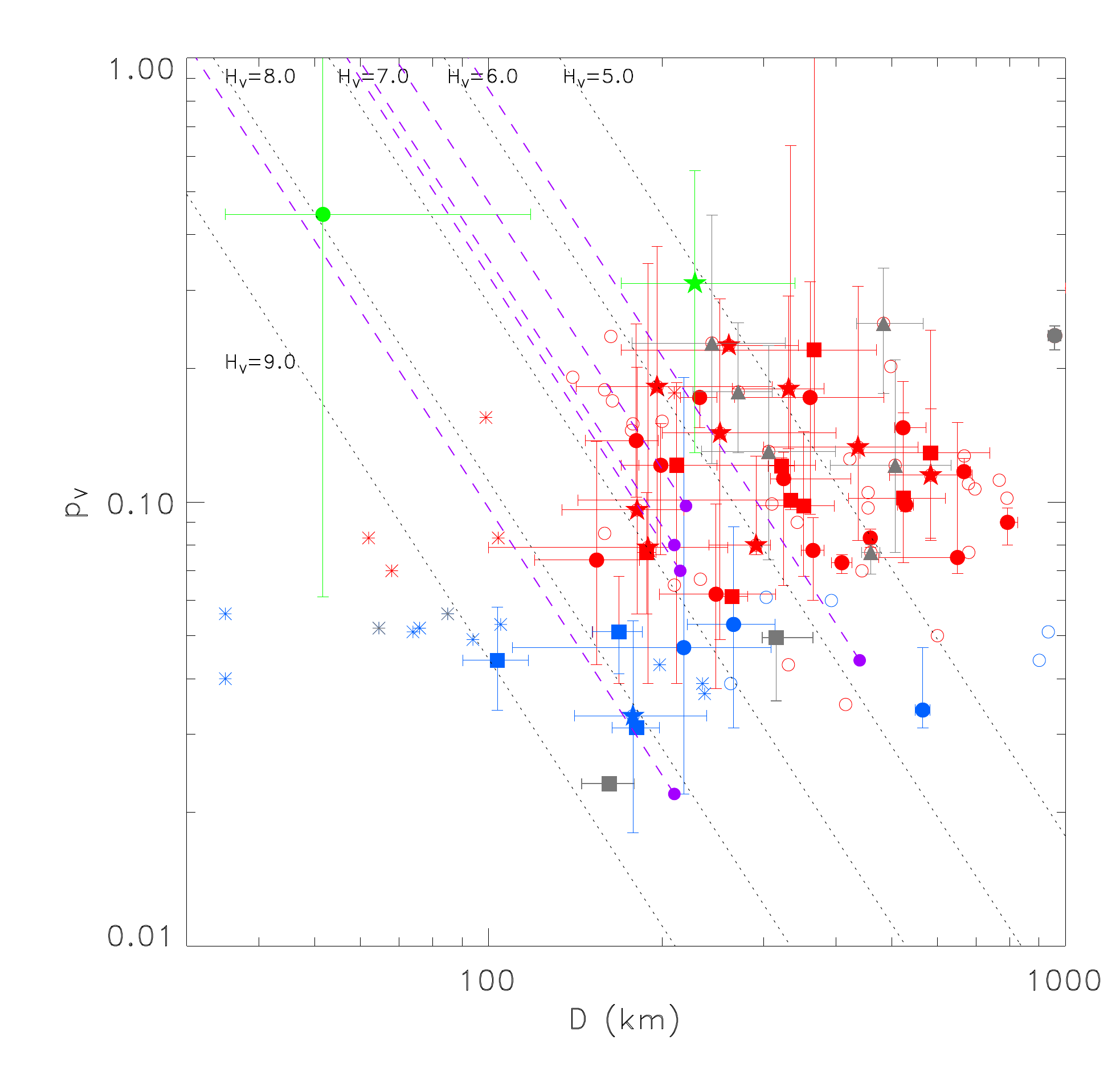}
\caption{Geometric albedo (\geomalb) vs. diameter (D) for the full scattered disk and resonant sample. The  clusters are color-coded: blue -- DN; red -- BR; green -- BN; gray -- ambiguous group assignment. Filled circles, stars, squares, and triangles indicate  plutinos, other resonances, scattered disk objects, and detached objects, respectively. Open circles are any other dynamical class, except the Centaurs that are shown as  asterisks. Dotted lines represent the constant absolute magnitudes, as  labeled. Purple symbols with dashed lines indicate those targets for which only diameter upper- and albedo lower limits could be derived in this paper.} 
\label{fig:diamvspv}
\end{figure}


It is also an important question whether objects with different sizes have the same typical albedos, and whether the same mean albedo (or albedo distribution) can be used to convert from absolute magnitudes to effective diameters within the same albedo-color group. In Fig.~\ref{fig:diamvspv} we plot the geometric albedos of resonant and scattered disk objects versus the effective diameter derived form radiometric data. We have to emphasis that this sample is not unbiased, as objects  in the ``TNOs are Cool'' studies (where the majority of these data are  from) were selected to cover different dynamic classes, sizes, or absolute magnitude ranges, and should have been foreseen to be detectable by the Herschel Space Observatory, assuming a uniform 8\% geometric albedo \citep{Muller2009,Kiss2014}. In the present sample no obvious trend is visible within any of the albedo-color groups. There is a general trend, considering all objects together, that smaller objects are darker, but this is dominated by the presence of Centaurs (asterisks in Fig.~\ref{fig:diamvspv}) that are found at smaller heliocentric distances and remain observable above \absmag\,$\lesssim$\,9.0, which is an absolute limit for the targets in the other dynamical families. In addition, darker objects are observable at smaller sizes in thermal emission at the same heliocentric distance, and that was the main criterion when the ``TNOs are Cool'' object list was compiled. This indicates that to our present knowledge a color-dependent and dynamical class-dependent albedo is probably suitable to transform between absolute magnitude and size distributions, as discussed above.

\section{Summary}

We determined the size and albedo of 23 trans-Neptunian resonant and scattered disk objects using data mainly based on the ``TNOs are Cool'' Herschel Open Time Key Programme observations, supplemented in some cases by Spitzer Space Telescope data. Together with the previous studies based on the ``TNOs are Cool'' data, this is
the largest reliable set of size and albedo estimates for objects in the outer resonances in the Kuiper belt. We confirm the results of a previous investigation \citep{Lacerda2014} that trans-Neptunian objects can be divided into albedo-color groups and that objects in relic populations exclusively have bright-red surfaces, strengthening the evidence for a compositional discontinuity in the young Solar System. We suggest that size distribution studies based on absolute magnitude distributions use the bimodal albedos, specific for the dark-neutral and bright-red groups in all dynamical populations, for example using    colors or spectral slopes as a proxy for albedo-color group membership \citep[e.g.,][]{Ayala}. 


\section*{Acknowledgements}
The research leading to these results has received funding from the European Union’s Horizon 2020 Research and Innovation Programme, under Grant Agreement no 687378; from the K-125015 and GINOP-2.3.2-15-2016-00003 grants of the National Research, Development and Innovation Office (NKFIH, Hungary). The work of G. Marton was supported by the NKFIH grant PD-128360. Part of this work was supported by the German DLR project number 50 OR 1108.
We thank our reviewer for the comments, especially those on the data reduction issues, which have certainly had an important impact on the final results of this paper. 




\newpage
\clearpage




\begin{thebibliography}{99}
%
\bibitem[Aaron et al.(2009)]{H1}
Aaron, F.D., Alexa, C., Anrdeev, V., et al. (the 'H1 collaboration'), 2009, Eur. Phys. J. C, 63, 625
\bibitem[Ali-Lagoa et al.(2018)]{AliLagoa2018}
Ali-Lagoa, V., M\"uller, T.G., Usui, F., Hasegawa, S., 2018, A\&A, 612, A85
\bibitem[Alvarez-Candal et al.(2011)]{AAC}
Alvarez-Candal, A., Pinilla-Alonso, N., Licandro, J., et al., 2011, A\&A, 532, A130
\bibitem[Ayala-Loera et al.(2018)]{Ayala}
Ayala-Loera, C., Alvarez-Candal, A., Ortiz, J.L., et al., 2018, MNRAS, 481, 1848
\bibitem[Bannister et al.(2018)]{Bannister}
Bannister, M.T., Gladman, B.J., Kavelaars, J.J., et al., 2018, ApJS, 236, 18
\bibitem[Barucci and Doressoundiram (1999)]{Barucci99}
Barucci, M.A., and Doressoundiram, A., 1999, Icarus, 142, 476
\bibitem[Boehnhardt et al.(2014)]{Boehnhardt}
Boehnhardt, H., Schulz, D., Protopapa, S., G\"otz, C., 2014,EM\&P, 114, 35
\bibitem[Bowell et al.(1989)]{Bowell1989}
Bowell, E.G., Hapke, B., Domingue, D., Lumme, K., Peltoniemi, J., Harris, A.W., 1989.
Application of photometric models to asteroids. In: Gehrels, T., Matthews, M.T.,
Binzel, R.P. (Eds.), Asteroids II. University of Arizona Press, pp. 524–555.
\bibitem[Brown \& Butler(2017)]{BB17}
Brown, M.E., Butler, B.J., 2018, AJ, 156, 164
\bibitem[Brucker et al.(2009)]{Brucker2009}
Brucker, M.J., Grundy, W.M., Stansberry, J.A., et al., 2009, Icarus, 201, 284
\bibitem[Buie et al.(2003)]{Buie2003}
Buie, M.W., Jordan, A.B., Wasserman, L.H., et al., 2003, MPEC 2003-H07
\bibitem[Delbo et al.(2003)]{Delbo2003}
Delbo, M., Harris, A.W., Binzel, R.P., Pravec, P., Davies, J.K., 2003, Icarus, 266, 116
\bibitem[Delbo et al.(2015)]{Delbo2015}
Delbo, M., Mueller, M., Emery, J. P., Rozitis, B., \& Capria, M. T. 2015, Asteroid Thermophysical Modeling, in: Asteroids IV., ed. P. Michel, F. E. DeMeo, \& W. F. Bottke, 107–128
\bibitem[DeMeo et al.(2009)]{DeMeo2009}
DeMeo, F.E., Fornasier, S., Barucci, M.A., 2009, A\&A, 493, 283 
\bibitem[DeMeo \& Carry(2014)]{DeMeo2014}
DeMeo, F.E. \& Carry, B., 2014, Nature, 505, 629
\bibitem[Doressoundiram et al.(2002)]{Dor2002}
Doressoundiram, A., Peixinho, N., de\,Bergh, C. et al., 2002, AJ, 124, 2279 
\bibitem[Doressoundiram et al.(2005)]{Dor2005}
Doressoundiram, A., Peixinho, N., Doucet, C., et al., 2005, Icarus, 174, 90
\bibitem[Doressoundiram et al.(2007)]{Dor2007}
 Doressoundiram, A., Peixinho, N., Moullet, A., et al., 2007, AJ, 134, 2186
\bibitem[Duffard et al.(2014)]{Duffard14}
Duffard, R., Pinilla-Alonso, N., Santos-Sanz, P., 2014, A\&A, 564, A92
\bibitem[Engelbracht et al.(2007)]{Engelbracht2007}
Engelbracht, C.W., Blaylock, M., Su, K.Y.L., 2007, PASP, 119, 994
\bibitem[Farkas-Tak\'acs et al.(2017)]{Farkas2017}
Farkas-Tak\'acs, A.I., Kiss, Cs., P\'al, A., et al., 2017, AJ, 154, 119
\bibitem[Fornasier et al.(2013)]{For13}
Fornasier, S., Lellouch, E., M\"uller, T.G., et al. 2013, A\&A, 555, A15
\bibitem[Fraley et al.(2018)]{Fraley}
Fraley, C., et al., 2018, Package 'mclust', version Novermber 17, 2018
\bibitem[Fraser et al.(2010)]{Fraser2010}
Fraser, W.C., Brown, M.E., Schwamb, M.E., 2010, Icarus, 210, 944
\bibitem[Fraser et al.(2014)]{Fraser2014}
Fraser, W.C., Brown, M.E., Morbidelli, A., Parker, A., Batygin, K., 2014, ApJ, 782, 100
\bibitem[Gladman et al.(2001)]{Gladman2001}
Gladman, B., Kavelaars, J. J., Petit, J.-M., et al., 2001, AJ, 122, 1051
\bibitem[Gladman et al. (2012)]{Gladman2012}
Gladman, B., Lawler, S.M., Petit, J.-M. et al. 2012, AJ, 144, 23
\bibitem[Gomes et al.(2008)]{Gomes2008}
Gomes, R. S., Fernández, J. A., Gallardo, T. et al. 2008, SSBN book, 259G
\bibitem[Gordon et al.(2005)]{Gordon2005}
Gordon, K.D., Rieke, G.H., Engelbracht, C.W., et al. 2005, PASP, 117, 503
\bibitem[Gordon et al.(2007)]{Gordon2007}
Gordon, K.D., Engelbracht, C.W., Fadda, D., et al. 2007, PASP, 119, 1019
\bibitem[Grundy et al.(2016)]{Grundy2016}
Grundy, W.M., Binzel, R.P., Buratti, B.J., et al., 2016, Science, 351, 9189 
\bibitem[Hainaut \& Delsanti(2002)]{Hai02}
Hainaut, O.R., \& Delsanti, A.C., 2002, A\&A, 389, 641
\bibitem[Hainaut \& Delsanti(2002)]{Hainaut2002}
Hainaut, O.R. \& Delsanti, A.C., 2002 A\&A 389, 641
\bibitem[Harris(1998)]{Harris98}
Harris, A.W. 1998, Icarus, 131, 291
\bibitem[Herschel Science Center(2009)]{HSC2009}
Herschel Science Centre, Herschel Products and Tools Contributor's Guide, HERSCHEL-HSC-DOC-1405, 4 December 2009
\bibitem[Horner \& Lykawka(2010)]{Horner2010}
Horner, J. \& Lykawka, P.S., 2010, MNRAS, 405, 49
\bibitem[Juri\'c et al.(2002)]{Juric2002}
Juri\'c, M., Ivezi\'c, \v{Z}, Lupton, R.H., et al., 2002, AJ, 124, 1776
\bibitem[Kiss et al.(2014)]{Kiss2014}
Kiss, Cs., M\"uller, Th.G., Vilenius, E., et al., 2014, ExA, 37, 161
\bibitem[Kiss et al.(2016)]{Kiss2016}
Kiss, Cs., P\'al, A., Farkas-Tak\'acs, A. I., et al., 2016, MNRAS, 457, 2908
\bibitem[Kiss et al.(2016b)]{Kiss2017}
Kiss, Cs., M\"uller, T.G., 2016, User Provided Data Product upload of Herschel/PACS near-Earth asteroid observations, Release Note V\,1.0,
available at: http://archives.esac.esa.int/hsa/legacy/UPDP/COOLTNOs/tnosarecool.pdf
\bibitem[Kiss et al.(2019)]{Kiss2019}
Kiss, Cs., Szak\'ats, R., Marton, G., et al., 2019, 'Small Bodies: Near and Far' database of thermal infrared measurements of small Solar System bodies, Release Note: Public Release 1.2, 2019 March 29 (http://ird.konkoly.hu)
\bibitem[Kiss et al.(2019b)]{Kiss2019b}
Kiss, C., Marton, G., Parker, A.H., et al., 2019, Icarus, 334, 3
\bibitem[Lacerda \& Jewitt(2007)]{Lacerda2007}
Lacerda P. \& Jewitt, D.C., 2007, AJ, 133, 1393
\bibitem[Lacerda(2011)]{Lacerda2011}
Lacerda, P., 2011, AJ, 142, 90
\bibitem[Lacerda et al.(2014)]{Lacerda2014}
Lacerda, P., Fornasier, S., Lellouch, E., et al., 2014,  ApJ, 793, L2
\bibitem[Lellouch et al.(2013)]{Lellouch2013}
Lellouch, E., Santos-Sanz P., Lacerda, P., et al., 2013, A\&A 557, A60
\bibitem[Lellouch et al.(2017)]{Lellouch2017}
Lellouch, E., Moreno, R., M\"uller, T., et al., A\&A 608, A45
\bibitem[Lellouch et al.(2016)]{L16}
Lellouch, E., Santos-Sanz, P., Fornasier, S., et al., 2016, A\&A 558, A2
\bibitem[Mommert et al.(2012)]{Mommert12}
Mommert, M., Harris, W., Kiss, Cs., et al., 2012, A\&A 541, A93
\bibitem[Mueller et al.(2012)]{Migo}
Mueller, M., Stansberry, J., Mommert, M. \& Grundy, W., 2012, 
"TNO Diameters And Albedos: The Final MIPS Dataset", AAS DPS meeting, \#44, \#310.13
\bibitem[M\"uller et al.(2009)]{Muller2009}
M\"uller, T.G., Lellouch, E., B\"ohnhardt, H., et al., 2009, EM\&P, 105, 209
\bibitem[M\"uller et al.(2010)]{Muller2010}
M\"uller, T.G., Lellouch, E., Stansberry, J., et al., 2010, A\&A, 518, L146
\bibitem[M\"uller et al.(2011)]{Muller2011}
M\"uller, T., Okumura, K., Klaas, U., 2011, PACS Photometer Passbands and Colour Correction Factors for Various Source SEDs, Herschel/PACS technical report, PICC-ME-TN-038

\bibitem[M\"uller et al.(2018)]{Muller2018}
M\"uller, T.G., Marciniak, A., Kiss, C., et al., 2018, Advances in Space Research, 62, 2326
\bibitem[M\"uller et al.(2018b)]{Muller2018b}
M\"uller, T. G., Kiss, C., Ali-Lagoa, V. et al., 2018, Icarus, in press (arXiv:1811.09476)
\bibitem[M\"uller et al.(2019)]{Muller2019}
M\"uller, T., Lellouch, E., Fornasier, S., 2019, TNOs and Centaurs at thermal wavelengths, https://arxiv.org/abs/1905.07158
\bibitem[Nielbock et al.(2013)]{Nielbock}
Nielbock, M., M\"uller, Th.G., Klaas, U., et al., 2013, Experimental Astronomy, 36, 631
\bibitem[Ortiz et al.(2017)]{Ortiz2017}
Ortiz, J.L., Santos-Sanz, P., Sicardy, B., et al., 2017, Nature, 550, 219 
\bibitem[PACS Data Reduction Guide (2015)]{PACS_guide}
PACS Data Reduction Guide: Photometry, Issue user. Version 12, Nov. 2015
\bibitem[PACS Observer's Manual(2011)]{PACS2011}
PACS Observer's Manual, HERSCHEL-HSC-DOC-0832, Version~2.3, 08-June-2011
\bibitem[P\'al et al.(2012)]{Pal12}
P\'al, A., Kiss, C., M\"uller, T.G., 2012, A\&A 541, L6
\bibitem[P\'al et al.(2016)]{Pal2016}
P\'al, A., Kiss, C., M\"uller, T.G., 2016, AJ, 151, 117
\bibitem[Perna et al.(2009)]{Perna2009}
Perna, D., Dotto, E., Barucci, M. A., et al., 2009, A\&A, 508, 451
\bibitem[Perna et al.(2010)]{Perna2010}
Perna, D., Barucci, M.A., Fornasier, S., et al., 2010, A\&A, 510, A53 
\bibitem[Perna et al.(2013)]{Perna2013}
Perna, D., Dotto, E., Barucci, M.A., et al., 2013, A\&A, 554, A49
\bibitem[Poglitsch et al.(2010)]{PACS}
Poglitsch, A., Waelkens, C., Geis, N., et al., 2010, A\&A, 518, L2
\bibitem[Romanishin \& Tegler(2005)]{Romanishin+Tegler}
Romanishin, W. \& Tegler, S.C., 2005, Icarus, 179, 523
\bibitem[Santos-Sanz et al.(2009)]{SS09}
Santos-Sanz, P., Ortiz, J.L., Barrera, L., Boehnhardt, H., 2009, A\&A, 494, 693
\bibitem[Santos-Sanz et al.(2012)]{SS12}
Santos-Sanz, P., Lellouch, E., Fornasier, S., 2012, A\&A, 541, A92
\bibitem[Schindler et al.(2017)]{Schindler2017}
Schindler, K., Wolf, J., Bardecker, J., et al., 2017, A\&A, 600, A12 
\bibitem[Sheppard(2012)]{Sheppard2012}
Sheppard, S.S., 2012, AJ, 144, 169
\bibitem[Sheppard \& Jewitt(2002)]{Sheppard02}
Sheppard, S.S., \& Jewitt, D.C. 2002, AJ, 124, 1757
\bibitem[Sheppard \& Jewitt(2004)]{Sheppard+Jewitt}
Sheppard, S.S. \& Jewitt, D.C., 2004, AJ, 127, 3023
\bibitem[Snodgrass et al.(2010)]{Snodgrass2010}
Snodgrass, C., Carry, B., Dumas, C., Hainaut, O., 2010, A\&A 511, A72
\bibitem[Spencer et al.(1989)]{Spencer1989}
Spencer, J.R., Lebofsky, L.A., Sykes, M.V., 1989, Icarus, 78, 337
\bibitem[Spencer(1990)]{Spencer1990}
Spencer, J.R., 1990, Icarus, 83, 27
\bibitem[Stansberry et al.(2007)]{Stansberry2007}
Stansberry, J.A., Gordon, K.D., Bhattacharya, B., 2007, PASP, 119, 1038
\bibitem[Stansberry et al.(2008)]{stansberry2008}
Stansberry, J., Grundy, W.M., Brown, M.E., et al., 2008, in The Solar System Beyond Neptune, 
Physical Properties of Kuiper Belt and Centaur Objects: Constraints from the Spitzer Space Telescope, p.161
(Tuscon, AZ: Univ. Arizona Press)
\bibitem[Stansberry et al.(2012)]{stansberry2012}
Stansberry, J.A., Grundy, W.M., Mueller, M., et al., 2012, Icarus, 219, 676
\bibitem[Stephens \& Noll(2006)]{SN2006}
Stephens, D.C. \& Noll, K.S., 2006, AJ, 131, 1142
\bibitem[Szak\'ats et al.(2020)]{Szakats2019}
Szak\'ats, R., M\"uller, T., Ali-Lagoa, V., et al., 2020, A\&A, 635, A54
\bibitem[Vilenius et al.(2012)]{Vilenius12}
Vilenius, E., Kiss, C., Mommert, M. et al., 2012, A\&A 541, A94
\bibitem[Vilenius et al.(2014)]{Vilenius14}
Vilenius, E., Kiss, C., Müller, T., 2014, A\&A 564, A35
\bibitem[Vilenius et al.(2018)]{Vilenius2018}
Vilenius, E., Stansberry, J., M\"uller, T.G., et al., 2018, A\&A, 618, A136
\bibitem[Volk et al.(2018)]{Volk}
Volk, K., Murray-Clay, Gladman, B.J., et al., 2018, AJ, 155, 260
\bibitem[Yu et al.(2018)]{Yu2018}
Yu, T.Y.M., Murray-Clay, R., Volk, K., 2018, AJ, 156, 33
\end{thebibliography}
\end{document}